\documentclass[a4paper,fleqn,usenatbib]{mnras}

\usepackage{newtxtext,newtxmath}
\usepackage[T1]{fontenc}
\usepackage{ae,aecompl}

\usepackage{graphicx}	% Including figure files
\usepackage{amsmath}	% Advanced maths commands
\usepackage{amssymb}	% Extra maths symbols

\title[Orbital Period Changes In Classical Novae]{Sudden and Steady Orbital Period Changes Across Six Classical Nova Eruptions; The End of Hibernation and Two Serious Challenges for the Magnetic Braking Model of Cataclysmic Variable Evolution}

\author[B. E. Schaefer et al.]{Bradley E. Schaefer$^{1}$\thanks{E-mail: schaefer@lsu.edu}
\\
$^{1}$Department of Physics and Astronomy, Louisiana State University, Baton Rouge, Louisiana, 70820, USA\\
}

\date{Accepted XXX. Received YYY; in original form ZZZ}

\pubyear{2019}

\begin{document}
\label{firstpage}
\pagerange{\pageref{firstpage}--\pageref{lastpage}}
\maketitle

\begin{abstract}
I report on two new measures of the sudden change in the orbital period ($P$) across the nova eruption ($\Delta$P) and the steady period change in quiescence ($\dot{P}$) for classical novae (CNe) RR Pic and HR Del, bringing a total of six such measures for CNe, all in a final report of my large and long observing program.  The fractional changes ($\Delta$P/P) in parts-per-million (ppm) are -290.71$\pm$0.28 (QZ Aur), -472.1$\pm$4.8 (HR Del), -4.46$\pm$0.03 (DQ Her), +39.6$\pm$0.5 (BT Mon), -2003.7$\pm$0.9 (RR Pic), and -273$\pm$61 (V1017 Sgr).  These results are in stark opposition to the Hibernation Model for the evolution of cataclysmic variables (CVs), which requires $\Delta$P/P$>$+1000 ppm to get the required drop in the accretion rate to produce hibernation.  The Hibernation Model cannot be salvaged in any way.  My program has also measured the first long-term $\dot{P}$ for classical novae, with -2.84$\pm$0.22 (QZ Aur), +4.0$\pm$0.9 (HR Del), +0.00$\pm$0.02 (DQ Her), -2.3$\pm$0.1 (BT Mon), and +1.25$\pm$0.01 (RR Pic) in units of $10^{-11}$ days/cycle.  These can be directly compared to the predictions of the Magnetic Braking Model, where the long-term average $\dot{P}$ is a single universal function of $P$.  The measured values are +5.3, $-$94, 0.00, +6.9, and $-$190 times that predicted by the model, so the predictions are always greatly wrong.  Further, the effects of the $\Delta$P averaged over the eruption cycle are usually much larger than the magnetic braking effects.  To get a realistic model of CV evolution, we must add the physics of the $\Delta$P and $\dot{P}$ variations.
\end{abstract}


%%%%%%%%%%%%%%%%%%%%%%%%%%%%%%%%%%%%%%%%%%%%%%%%%%

%%%%%%%%%%%%%%%%% BODY OF PAPER %%%%%%%%%%%%%%%%%%

\section{Introduction}

Cataclysmic variables (CVs) are semidetached binaries where a normal star has Roche lobe overflow of material onto a white dwarf (WD).  The most prominent examples are classical novae (CNe) and recurrent novae (RNe), with their spectacular thermonuclear eruptions on a recurrence time scale of $\tau_{rec}$.  By massive observing and theory work over the last half-century, we now largely understand CVs and CNe in detail (e.g., Bode \& Evans 1989; 2008).  Now, as always, there are questions and controversies relating to details of individual properties, mechanisms, and interpretation; and these are important.  But the big-picture questions now at the forefront of our field are related to the {\it evolution} of CVs.  The evolution of CVs is what allows us to relate the myriad CV classes to each other, to show and explain their commonalities and differences, and to understand CV histories and futures.  It is evolutionary studies that explain the life cycles of CVs.  (A similar situation is for biologists, where detailed studies of lions, tigers, and bears are important, but it is evolutionary studies that explain why there are lions and tigers and bears, as well as giving the big picture of how they all relate.)  Evolution is needed to understand the big picture of CVs.

Various CV evolution schemes has been proposed, although little is now confident.  The two most prominent and venerable schemes are the Hibernation model (Shara et al. 1986; Prialnik \& Shara 1986; Livio \& Shara 1987; Shara 1989) and the Magnetic Braking Model (MBM; Rappaport, Joss \& Webbink 1982, Patterson 1984, Knigge, Baraffe \& Patterson 2011).  The Hibernation model describes the cyclic path of a single system from a CN, to a dwarf nova, to a quiescent disconnected binary, to a dwarf nova, to a CN, and so on, as the accretion rate changes.  The time scale for each cycle is perhaps a million years, and it is driven entirely by the disconnect caused by the sudden change in the orbital period caused by the mass lost during each nova eruption.  The MBM quantifies the theoretical idea that the orbital periods of all CVs must be grinding down due to various mechanisms that lose angular momentum from the binary orbit, and this period change over billions of years will drive what the system looks like.  Thus, after a binary comes into connection, it starts out with a long orbital period at a high accretion rate, and steadily follows a prescribed path to shorter orbital periods at lower accretion rates.

Unfortunately, for observational study of CV evolution, we mostly do not have any measures over long-time spans, so evolutionary effects are not visible.  That is, with modern CCDs and space telescopes, our community has at most a decade or two of history for any one CV, and this forces us into the always-uncertain and controversial game of piecing together an evolutionary progression by pasting together snapshots of individual stars.  (A similar problem would be to try to understand the movie ``Gone With the Wind" from a scattered pile of single-frame outtakes mixed in with single-frame outtakes from other movies.)  An effective partial solution to this deep problem is to use archival data to get a history of many CVs for the last 130 years.  With this, we can measure the time derivatives of properties, and such help enormously for piecing together snapshots.  For example, the critical component of both the Hibernation model and MBM is the {\it changes} in the orbital period, and this can only be measured with long stretches of archival data.  (Imagine for the ``Gone With the Wind" problem, having many short film clips would help tremendously in working out the progress of the film and plot.)  A history lasting 130 years is long enough that the evolutionary effects will become detectable above the chaotic variability.

The historical record from archival data is largely in the form for which we can pull out photometric information, i.e., we only get light curves.  For most CVs, the record starts with the Harvard sky photographs (plates) which start in 1889.  For every star in the sky (both north and south) brighter than 17th mag or so, the Harvard plates will yield hundreds-to-thousands modern Johnson B magnitudes from 1889-1989.  This can be supplemented with modern archival magnitudes to bring the light curve to the current date.  For CVs, with 130 years of photometry, we can chart out the accretion rate changes, the orbital period changes, and the changes from the nova eruptions.  These are exactly the most important properties to describe and interpret evolutionary changes in CVs.  These are obtainable only with long runs of archival data.

Archival data can pull out evolutionary effects in a variety of ways, all of which are based only on simple light curves covering a century or so:  Long after the nova eruption has stopped, does the old nova keep fading away, heading towards a hibernation state as predicted and required by the Hibernation model (e.g., Collazzi et al. 2009; Johnson et al. 2014)?  Does the orbital period ($P$) change across a nova eruption ($\Delta$P), from a pre-eruption orbital period ($P_{pre}$) to a post-eruption orbital period ($P_{post}$), increase by the large amount as predicted and required by the Hibernation model (e.g., Salazar et al. 2017; Schaefer et al. 2019)?  Does the steady orbital period change during quiescence ($\dot{P}$)  decrease in the manner as predicted and required by the MBM (e.g., Schaefer et al. 2019)?

I have been pursuing a program to measure orbital period and brightness changes of CNe and RNe.  This program has taken roughly 300 nights of my own telescope time, plus dozens of trip to observatories around the world with archival data.  I started this in 1983 and have been working on this continuously since.  From the beginning, I knew that major parts of this program would take several decades to complete.

This paper has a companion paper.  The companion paper includes:  (1) A detailed discussion of the physics and equations for both sudden and steady orbital period changes in nova systems.  (2) A detailed description, motivations, and overview of my whole $\Delta$P program.   (3) A detailed description of the capabilities and techniques for photometry with archival photographic plates. (4) A report on my newly measured $\Delta$P and $\dot{P}$ for two CNe DQ Her and BT Mon.  (5) A brief analysis of the implications of the four CNe with measured $\Delta$P.

This paper starts with reporting my newly discovered $P_{pre}$ values for RR Pic (Section 2) and HR Del (Section 3), from which I use archival data to also get their $P_{post}$ values and hence $\Delta$P values.  Necessarily, I measure $\dot{P}$ as a by-product.  In Section 4, I present a full summary and overview of my measured results for all six CNe, with the overview table being the primary observational result of my entire $\Delta$P program for CNe.  Section 5 is a direct testing of the Hibernation model predictions with my $\Delta$P results.  Section 6 is a direct testing of the MBM predictions with my period change results.  Section 7 gives a summary of the rather far-reaching conclusions for CV evolution.

This paper is the final paper reporting on the completion of my long-running program to measure $\Delta$P for CNe.  I have made complete use of archival data for all the CNe that can possibly have a measured $\Delta$P.  There are no more useable archival data for the six CNe, and no more CNe for which we can possibly get $P_{pre}$, so the six $\Delta$P values are all that we can have for a long time to come.

\section{RR Pic}

RR Pic (Strope, Schaefer \& Henden 2010) peaked at magnitude 1.0, making it the third all-time brightest nova (behind V603 Aql and GK Per).  The eruption started on JD 2424298 (1925.402), and peaked 11 days later.  The time to decline by three magnitudes from peak (i.e., $t_3$) was 122 days.  The light curve displayed three sharp peaks (`jitters') and is therefore of the relatively low-energy J-class.  The decline was very slow, getting back to the pre-eruption quiescent level only around 1970.  RR Pic has an observed expanding nova shell, while its WD is magnetic.  In quiescence, the average magnitude is V=12.2, making it the fourth brightest known classical nova (behind V603 Aql, HR Del, and V2491 Cyg).  The system in quiescence shows superhumps and quasi-periodic oscillations (Schmidtobreick et al. 2008).  For a distance of 511$\pm$8 pc from {\it Gaia} DR2, the absolute magnitude in quiescence is $M_V$=3.7 (Schaefer 2018).  Van Houten (1966) discovered the orbital modulation of RR Pic with a period of 0.1451 days (3.48 hours).  The presence of substantial orbital modulations and the bright counterpart means that is might be possible to measure a $P_{pre}$ value from archival plates 1890--1925, and hence to derive a $\Delta$P.

\subsection{RR Pic Post-Eruption Orbital Period}

\begin{table}
	\centering
	\caption{New Times of Maximum Light for RR Pic}
	\begin{tabular}{llll} 
		\hline
		Data Source	&   Year   & $T_{max}$ (HJD) & $O-C$ (days)  \\
		\hline
HCO (1889--1905)	&	1899.63	&	2414886.0199	$\pm$	0.0087	&	...	\\
HCO (1905--1925)	&	1915.69	&	2420751.0200	$\pm$	0.0054	&	...	\\
HCO (1944, 1945)	&	1945.24	&	2431542.1396	$\pm$	0.0022	&	0.1151	\\
HCO (1946, 1947)	&	1946.93	&	2432160.0761	$\pm$	0.0033	&	0.0960	\\
HCO (1948, 1949)	&	1948.89	&	2432878.0792	$\pm$	0.0092	&	0.0756	\\
HCO (1950--1952)	&	1951.28	&	2433751.1321	$\pm$	0.0077	&	0.0722	\\
AAVSO (MGW)	&	2017.87	&	2458071.0488	$\pm$	0.0002	&	0.0057	\\
AAVSO (NLX)	&	2017.87	&	2458072.0603	$\pm$	0.0012	&	0.0020	\\
AAVSO (HMB)	&	2017.93	&	2458095.1235	$\pm$	0.0009	&	0.0061	\\
		\hline
	\end{tabular}
\end{table}

	To derive a $\Delta$P, we need a $P_{post}$.  Various reported $P_{post}$ values are 0.1451 days (Van Houten 1966), 0.1450255$\pm$0.0000002 days (Vogt 1975), 0.14502545$\pm$0.00000007 days (Kubiak 1984), and 0.145025959$\pm$0.000000015 days (Vogt et al. 2017).
		
	The orbital modulation is roughly a quarter of a magnitude, with the maximum being fairly broad and flat.  The folded light curve shape shows considerable structure and variability from year-to-year, including up to three separate peaks in the broad maximum (Vogt 1975) and possible grazing eclipses (Warner 1986).  In all cases, the light curves show the usual flickering at the 0.2 magnitude level.  This all makes for it being difficult to define or measure some fiducial orbital phase.  Unfortunately, for times of reported maxima from the literature, we never have a clear definition (and no original data to allow a consistent analysis), so there is a real possibility of substantial systematic error from inconsistencies in the definitions of times of maxima for each source.  This will make for increased scatter in the $O-C$ curve, while the source-by-source nature makes it difficult to interpret small changes in the $O-C$ curve.  Further, the noise in times caused by the usual flickering, the quasi-periodic oscillations, and the superhumps makes for the real total uncertainty being substantially larger than the quoted measurement error bars.  This means that the measurement errors are negligible, and the real uncertainties in the $O-C$ values can only come by observing the scatter of measured points.  Nevertheless, these effects are small enough as to be largely negligible for determining the post-eruption orbital period in the last 54 years to all needed accuracy.
	
	Vogt et al. (2017) reports an excellent analysis of 203 epochs of maximum light as observed by themselves and as collected from the literature.  Their basic linear fit in the $O-C$ curve gives a period of 0.145025959$\pm$0.000000015 days.  Their $O-C$ plot is given in Fig. 1, constituting all the points from 1965 to 2014.  Their values show significant differences from the best-fit line with a time scale of decades.  They tried fitting these deviations with a parabola and with a third-body orbit.  None of these fits is convincing, for example their $O-C$ covers 49 years, which is less than three-quarters of their third-body orbital period.  Their observations cannot distinguish between the possibilities.  The extrapolation of their plausible models back to 1925.402 produces an uncertainty in the epoch of maximum light that is larger than an orbital period.
	
\begin{figure}
	\includegraphics[width=\columnwidth]{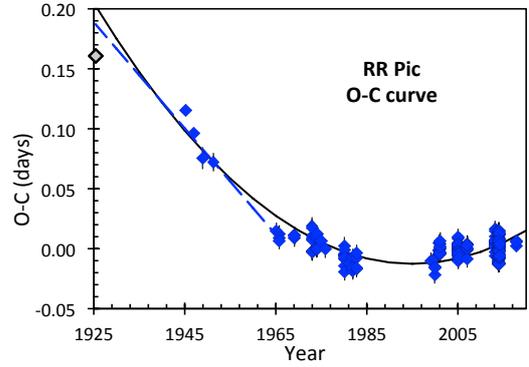}
    \caption{$O-C$ curve for RR Pic.  The $O-C$ is the deviation in the times of maximum light versus the linear ephemeris of Vogt et al. (2017).  The 203 values measured-by and collect-by Vogt et al. (2017) are represented as blue diamonds from 1965--2014, with the average error bars displayed.  Here, I add four more points for 1944--1952 as based on Harvard plates plus three points for 2017 as based on CCD time series recorded in the AAVSO data archives.  The best fit parabola reasonably describes the variations, implying a steady period change with $\dot{P}$=(1.4$\pm$0.2)$\times$10$^{-11}$ days per cycle, although there are significant, but small, deviations from this parabola.  The slope of the extrapolated parabola in 1925.402 (the start of the nova event) gives $P_{post}$.  Alternatively, a linear extrapolation from the 1944--1965 leads to a slightly different $P_{post}$.  The epoch of maximum light close to the start of the eruption has an $O-C$ corresponding to near when the extrapolations cross the left axis.  The extrapolation of the pre-eruption best fit yields an epoch of maximum light corresponding to the large gray diamond near the left axis.  The utility of this figure is to show convincingly that the overall shape of the $O-C$ curve is closely parabolic, to show how $P_{post}$ is determined, and to show that the extrapolation has substantial uncertainty.}
\end{figure}
	
	I can improve on this work of Vogt et al. (2017) by adding epochs from 1944--1952 and 2017.  These extensions give a length for the $O-C$ curve from 1944--2017 (74 years) as an improvement over the prior work covering 1965--2014 (49 years).  The 2017 data come from the AAVSO database, where three observers archived long time series of CCD magnitudes on many nights in the V-band.  The 1944--1952 data come from the Harvard plates, showing the nova late in its tail after the large amplitude sinewave variations started up.  The times of maximum light were derived by fitting sinewaves to the light curve in a chi-square sense, and the time of maximum light was taken to be the time of the peak of the best-fit sinewave.  I chose a peak time close to the average times of the input observations.  The heliocentric Julian dates for each of these seven added epochs of maximum light are presented in Table 1, and the $O-C$ values (based on the linear ephemeris of Vogt et al. 2017) are added in Fig. 1.
	
	The 2017 V-band CCD time series recorded in the AAVSO database are for three observers; Franz-Josef Hambsch (from Belgium), Gordon Myers (from the US), and Peter Nelson (from Australia), with observer codes of HMB, MGW, and NLX respectively.  The light curves include 330, 6362, and 510 individual magnitude measures for the three observers respectively.  Hambsch has already provided a huge amount of excellent data for the Vogt et al. (2017) paper.  The folded light curve for the Meyers data is shown in Fig. 2.  Note the usual large scatter around the average light curve due to RR Pic displaying strong flickering, superhumps, quasi-periodic oscillations, and some secular variations.
	
\begin{figure}
	\includegraphics[width=\columnwidth]{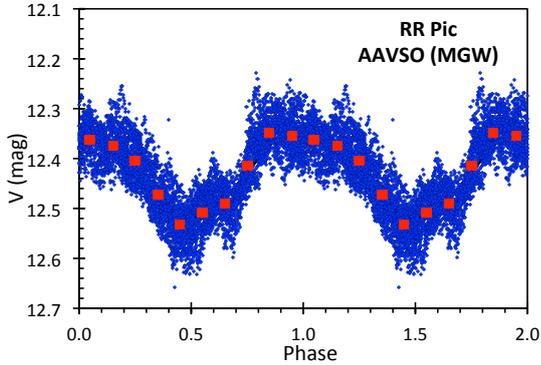}
    \caption{Folded light curve for RR Pic from the Myers data.  These 6362 V-band CCD measures are depicted as small blue diamonds, while the phase-averaged folded light curve is shown as large red squares.  The best fit sinewave is a black curve underneath the blue diamonds, and it is largely invisible, which makes the point that the light curve is roughly a sinewave.  The epoch of maximum light is taken from the best fit sinewave, as given in Table 1.}
\end{figure}
	
	The 1944--1952 B-band light curve (Fig. 3) is from 347 Harvard plates.  This is late in the tail of the fading nova eruption, where the light from the inner binary star dominates over nebular light.  Before 1944, the total nova light is so bright so as to dominate the periodic variations of the inner binary system, so that no significant periodicity is seen in the light curve.  After 1952, the infamous Menzel Gap stopped the Harvard plates, so the new $O-C$ curve has a gap until photoelectric photometry started up in 1965.  The best fit period over this interval is 0.1450221$\pm$0.0000006 days.  This is significantly different from all the other published periods for later epochs.  I have broken this Harvard data into four time intervals so as to derive four epochs of maximum light, as these can then be plotted on an $O-C$ diagram.
	
\begin{figure}
	\includegraphics[width=\columnwidth]{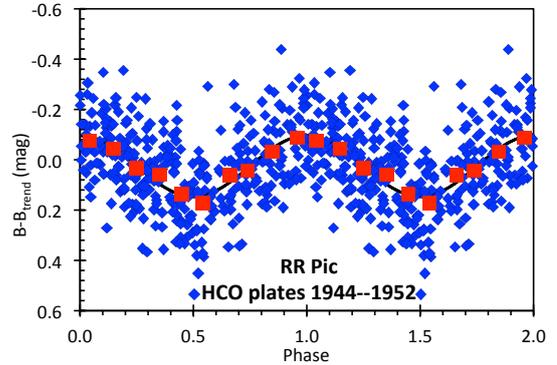}
    \caption{Folded light curve for RR Pic from 347 Harvard plates 1944-1952.  These B-band magnitudes (blue diamonds) are from late in the tail of the 1967 eruption, with a smooth trend curve being subtracted out.  The phase-binned light is shown as large red squares, and the best fit sinewave is the black curve.  We see the periodic modulation with a large amount of scatter.  This scatter is mostly from intrinsic variations in RR Pic (flickering, superhumps, and quasi-periodic oscillations), while the measurement errors are comparatively small.}
\end{figure}
	
	The $O-C$ diagram over 1944--2017 shows an approximately parabolic shape.  (Note, it is possible to move the 1944-1952 points up or down by one orbital period, assuming some error from the obvious cycle count.  Such real possibilities would require at least two large and sudden changes of period between 1952 and 1965, so Occam's Razor strongly points to the simple and direct answer as shown.)  The best fit parabolic model is displayed as a black parabola.  The steady period change has $\dot{P}$=(1.4$\pm$0.2)$\times$10$^{-11}$ days per cycle.  With this, during quiescence, the orbital period is steadily {\it increasing}, so the binary separation must be getting steadily larger and larger over time.  We see significant deviations from the parabola.  I judge that these deviations are larger than is possible by various observers having different definitions of maximum light.  The deviations are much larger than the scatter in the $O-C$ curve, so that means that ordinary variations from flickering, superhumps, and secular variations cannot account for the deviations.  So I can only think that RR Pic has some additional mechanism for small period changes superposed on top of some steady mechanism that give a constant $\dot{P}$.
	
	For purposes of this paper, to get $\Delta$P, we need $P_{post}$ at the time of the eruption, and this can only come from the post-eruption data.  Further, it would be helpful to get an epoch of maximum light at the time of the eruption, as this can help with the measure of $P_{pre}$ with a joint fit to before and after the nova.  Unfortunately, the $O-C$ curve cannot be extrapolated back to 1925.402 with high accuracy.  For example, Fig. 1 shows the best overall parabolic fit extrapolated to 1925.402, as well as the best straight line fit to the 1944--1965 data also extrapolated back to the year of the nova.  The two lines cross the date of the eruption with substantial difference in $O-C$, and we all can see other extrapolation schemes that would yield even larger differences.  The extrapolated $O-C$ can be represented by 0.20$\pm$0.02 days, but the real uncertainty could well be larger.  And the indicated $P_{post}$ (i.e., the slope of these extrapolations in 1925.402) has substantial uncertainty.  Fortunately, the uncertainties in the extrapolated $P_{post}$ values are greatly smaller than the $\Delta$P.
	
	Still, we need an estimate of $P_{post}$ for the date of the nova.  The best-fit parabola has $P_{post}$=0.1450235 days.  I will adopt this value as the best $P_{post}$.  The uncertainty depends on the extrapolation back to 1925.402, with many plausible schemes.  The extrapolation of the 1944--1965 line has $P_{post}$=0.1450242 days.  For linear extrapolation ranges from 1944--1949 to 1944--1982, half the variation in $P_{post}$ is 0.0000018 days.  So I am concluding that $P_{post}$ is close to 0.1450235$\pm$0.0000018 days.  These numbers will be substantially improved with a joint fit of both pre-eruption and post-eruption data, because the $O-C$ curve must be continuous across the eruption.

\subsection{RR Pic Pre-Eruption Magnitudes}

The only way to get $P_{pre}$ is to measure the period from the sinusoidal oscillations in the pre-1925 light curve.  The only way to get a pre-eruption light curve is to use the many Harvard plates from 1889--1925.402.  (No other set of archival plates has coverage before 1925 with the needed depth.)  Fortunately, RR Pic is bright so as to appear on many plates, and fortunately, the sinewave must have something like a quarter-magnitude amplitude.  So it should be easy to get $P_{pre}$.

	To this end, I have examined all relevant Harvard plates on four different occasions.  These multiple measures of magnitudes were to provide tests of measurement accuracy, and to beat down measurement errors.  All measures were done completely independently, with no knowledge of which plate was being measured nor any prior measurement results.  I found that all measures were consistent.  (Work on other stars also found consistency with the DASCH measures, where the visual measures have equal error bars as those given by DASCH.)  The average RMS for these multiple measures of the same plate is 0.08 mag.  This is the measurement error from the plate.
	
	The total photometric error will be the addition in quadrature of the measurement errors (averaging to 0.08 mag in the case of RR Pic) and the random scatter due to vagaries in the plate material, including photon statistics in the development of grains.  For a bright star like RR Pic, these extra errors will be small.
	
	The scatter in the light curve will be from the addition in quadrature of the photometric errors plus the instance-to-instance variations in the intrinsic brightness of RR Pic.  Unlike most eclipsing binary stars with effectively zero variation from orbit-to-orbit, RR Pic suffers large amplitude variations due to strong flickering, superhumps, quasi-periodic oscillations, and secular variations.  So when a folded light curve is constructed, the individual points will show a wide scatter around any single light curve shape.  This is seen in Fig. 2 and 3.  This is a perpetual problem for period work on all novae and cataclysmic variables.  For display purposes, it is usually more revealing to show an averaged light curve, with the averages running over some set of phase bins.  
	
	The best measure of the total error in reproducing some light curve model comes from the RMS scatter about the model.  For the best fit sinewave (see Section 2.4), the RMS scatter in each phase bin  is 0.19 mag.  This includes both photometric error plus the random variability of RR Pic itself.  The intrinsic variability dominates over the photometric errors, and this is applicable to all data points.  This also implies that we could not have done better if the photometric accuracy of the Harvard plates were greatly reduced.  That is, even CCD measures (instead of photographic plates) would not provide a significant improvement.
	
	Each plate was measured up to four times, with the measured magnitude then averaged.  The plate material has a native magnitude system very close to the Johnson B band, so there are no significant color terms.  The comparison stars used were those provided by the APASS survey, and hence are in the B band.  So the resultant magnitudes are in the Johnson B-band, consistently throughout the history of the Harvard blue plates.  The times for the middle of each exposure are corrected to HJD.  The result is 82 Johnson B magnitudes from 1889 to 1925.1, each with a total scatter (from photometric errors and from intrinsic variations) of $\pm$0.19 mag, as displayed in Table 2.

\begin{table}
	\centering
	\caption{Pre-eruption light curve for RR Pic}
	\begin{tabular}{llll} 
		\hline
		$T_{mid}$ (HJD)	&   Year   & B& HCO plate  \\
		\hline
2411297.8737	&	1889.807	&	12.89	&	B  4497	\\
2411612.8740	&	1890.670	&	12.93	&	B  5640	\\
2411612.8820	&	1890.670	&	12.80	&	B  5641	\\
2411612.8900	&	1890.670	&	13.20	&	B  5642	\\
2412936.5134	&	1894.293	&	12.87	&	B 10914	\\
2413153.8502	&	1894.889	&	12.82	&	B 12642	\\
2413209.5530	&	1895.041	&	12.98	&	B 12779	\\
2413491.7667	&	1895.814	&	12.90	&	B 14950	\\
2413495.8068	&	1895.825	&	13.12	&	B 14983	\\
2413888.7153	&	1896.899	&	13.06	&	B 18006	\\
2413892.7784	&	1896.911	&	12.85	&	B 18110	\\
2414169.9040	&	1897.670	&	12.80	&	B 20235	\\
2414277.6787	&	1897.965	&	13.14	&	B 20912	\\
2414613.8012	&	1898.886	&	12.97	&	B 22257	\\
2414766.5053	&	1899.304	&	13.33	&	B 22597	\\
2414921.8622	&	1899.730	&	13.28	&	B 24195	\\
2414940.8645	&	1899.782	&	13.07	&	B 24370	\\
2414942.7795	&	1899.787	&	13.02	&	B 24430	\\
2415071.6411	&	1900.140	&	13.15	&	A 4206	\\
2415292.8853	&	1900.746	&	13.08	&	B 26338	\\
2415318.8507	&	1900.817	&	12.95	&	B 26497	\\
2415334.7725	&	1900.861	&	13.00	&	AM  712	\\
2415340.7701	&	1900.877	&	13.12	&	B 26608	\\
2415380.7548	&	1900.987	&	13.10	&	B 26832	\\
2415415.6141	&	1901.083	&	13.00	&	B 26926	\\
2415779.6421	&	1902.080	&	13.21	&	B 29113	\\
2415841.5373	&	1902.249	&	13.08	&	A 5772	\\
2415865.5652	&	1902.315	&	12.67	&	B 29417	\\
2415866.5211	&	1902.318	&	13.00	&	A 5817	\\
2416034.8415	&	1902.779	&	13.22	&	B 30903	\\
2416130.6438	&	1903.041	&	12.60	&	AM  1797	\\
2416131.6780	&	1903.044	&	13.42	&	A 6291	\\
2416179.5891	&	1903.175	&	12.90	&	AM  1830	\\
2416230.4842	&	1903.315	&	12.91	&	B 31531	\\
2416240.4971	&	1903.342	&	13.10	&	B 31614	\\
2416359.8766	&	1903.670	&	12.80	&	AM  2250	\\
2416466.6877	&	1903.962	&	13.04	&	B 32931	\\
2416583.5385	&	1904.282	&	13.10	&	B 33301	\\
2416726.8791	&	1904.673	&	13.20	&	AM  3009	\\
2416787.7459	&	1904.839	&	12.80	&	B 35148	\\
2417094.8751	&	1905.680	&	13.20	&	AM  3848	\\
2417105.8275	&	1905.710	&	13.10	&	AM  3888	\\
2417110.8177	&	1905.724	&	13.20	&	AM  3909	\\
2417128.8741	&	1905.774	&	13.20	&	AM  3951	\\
2417212.6579	&	1906.003	&	12.73	&	A 7558	\\
2417213.6047	&	1906.006	&	12.70	&	AM  4040	\\
2417298.5279	&	1906.238	&	12.65	&	AM  4151	\\
2417465.8747	&	1906.697	&	13.00	&	AM  4569	\\
2417524.7515	&	1906.858	&	13.00	&	AM  4675	\\
2417630.6415	&	1907.148	&	12.95	&	A 8217	\\
2418027.5657	&	1908.235	&	12.94	&	B 38486	\\
2418042.4465	&	1908.276	&	13.40	&	AM  5387	\\
2418279.8704	&	1908.925	&	13.19	&	B 39844	\\
2418614.4494	&	1909.900	&	13.05	&	AK 569	\\
2418614.8626	&	1909.842	&	13.20	&	AM  6746	\\
2418626.4047	&	1909.900	&	13.15	&	AK 577	\\
2418647.8054	&	1909.932	&	13.20	&	AM  6804	\\
2418997.8282	&	1910.891	&	12.67	&	B 42129	\\
2419029.7588	&	1910.979	&	13.13	&	A 10356	\\
2420421.7698	&	1914.789	&	12.80	&	AM  10196	\\
2420490.7463	&	1914.979	&	12.50	&	AM 10317	\\
2420611.5193	&	1915.310	&	12.65	&	AM 10474	\\
2420769.8168	&	1915.743	&	12.80	&	AM 11284	\\
2420814.8173	&	1915.867	&	12.50	&	AM 11448	\\
		\hline
	\end{tabular}
\end{table}

\begin{table}
	\centering
	\contcaption{Pre-eruption light curve for RR Pic}
	\label{tab:continued}
	\begin{tabular}{lllll} 
		\hline
		$T_{mid}$ (HJD)	&   Year   & B& HCO plate  \\
		\hline
2421194.7634	&	1916.905	&	12.65	&	AM 12800	\\
2422000.5935	&	1919.112	&	12.91	&	MF  3130	\\
2423353.8542	&	1922.817	&	12.95	&	AM 16129	\\
2423706.8597	&	1923.785	&	13.15	&	AM 16445	\\
2423768.6677	&	1923.954	&	13.20	&	AM 16477	\\
2423780.7104	&	1923.986	&	13.07	&	MF  8264	\\
2423780.7394	&	1923.986	&	13.17	&	MF  8265	\\
2423791.6661	&	1924.017	&	13.10	&	AM 16531	\\
2423798.6605	&	1924.035	&	12.81	&	MF  8336	\\
2423798.6745	&	1924.035	&	12.87	&	MF  8335	\\
2423800.6209	&	1924.041	&	13.10	&	AM 16572	\\
2423828.5336	&	1924.116	&	13.13	&	MF 8412	\\
2423828.5411	&	1924.118	&	13.20	&	MF  8412	\\
2424084.7714	&	1924.750	&	13.10	&	AX 888	\\
2424122.7876	&	1924.922	&	12.99	&	MF 8957	\\
2424122.8119	&	1924.922	&	12.98	&	MF 8959	\\
2424144.6154	&	1924.981	&	13.20	&	AM 16681	\\
2424200.6236	&	1925.135	&	12.22	&	MF 8991	\\
		\hline
	\end{tabular}
\end{table}

\subsection{RR Pic Pre-Eruption Orbital Period}

The 82 pre-eruption magnitudes were run through a discrete Fourier transform (DFT) to look for peaks.   The range over which periodicities was sought extended from 0.144589 to 0.145459 days, as being within 3000 ppm of $P_{post}$.  Physical limitations (e.g., in the asymmetric ejection of the nova shell, see Schaefer et al. 2019) restrict the period change to be less than this search limit.  Since we are strongly expecting the pre-nova system to display sinusoidal modulations with roughly a quarter of a magnitude amplitude (i.e., just like the current quiescent nova), there really must be a significant signal in the DFT.  The DFT is shown in Fig. 4.  And, we do see one peak standing greatly above the ordinary noise peaks.  This is at a period of $P_{pre}$=0.14531361 days.  So we already know the answer as to the value of $P_{pre}$.
	
\begin{figure}
	\includegraphics[width=\columnwidth]{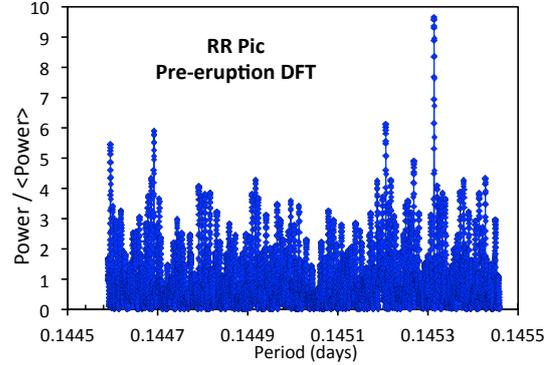}
    \caption{Discrete Fourier transform of 82 pre-eruption magnitudes for RR Pic.  The highly oversampled DFT covers a range of trial periods within 3000 ppm of $P_{post}$.  We see one peak that sticks out far above the ordinary noise spikes.  (The noise spikes over a very broad range never get above 6.6, so the high peak in this range is highly significant.)  This peak is at a period of 0.14531361 days, and the sinewave shaped folded light curve has a peak-to-peak amplitude of 0.276 mag.  As the pre-eruption orbital modulation must be something like a quarter-magnitude, it must have a peak in the DFT much like what is seen here.}
\end{figure}

	The DFT is optimal for the discovery of nearly-sinusoidal periodicities, but it does not readily provide $E_0$ or amplitude values, nor allow for any $\dot{P}$, nor allow significance calculations, nor offer any means to determine error bars.  To solve these problems, I use the standard chi-square fitting of the 82 magnitudes in the light curve to a simple sinewave.  This provides a well-known and easy method to derive the various model parameters and their uncertainties.  For the chi-square analysis, the fit parameters are $P_{pre}$, $E_0$, the amplitude, and $\dot{P}$.  
	
We would expect that the $\dot{P}$ is the same during the pre-eruption and post-eruption quiescence.  The pre-eruption light curve does not have either the number of points or the years of coverage to make a significant detection of any reasonable $\dot{P}$ term.  This is as expected, because we have only 82 pre-eruption magnitudes extending over 36 years.  For comparison, with many tens of thousands of individual magnitudes (leading to 203 accurate measures of times of maximum brightnesses), spread over 49 years, Vogt et al. (2017) was not able to make any confident discovery of the parabolic term.  In detail, my chi-square fits have no difference in the best fit parameters between the cases where $\dot{P}$ varies from zero to its post-eruption value.  The period at the end of the pre-eruption interval changes only slightly (far smaller than the quoted error bars) between these two cases.  For this result, we are further helped by the lack of any need to extrapolate, as we have the pre-eruption light curve up to three months before the eruption.  So in all, the effects of changing $\dot{P}$ is completely negligible for the pre-eruption fits.
	
	The best fit returns 0.14531361$\pm$0.00000019 d for an epoch of HJD 2417818.0097$\pm$0.0051 and a full peak-to-peak amplitude of 0.276$\pm$0.058 mag.  The epoch of maximum light at the time of the eruption is 2424298.1248$\pm$0.0084 for the $\dot{P}$=0 case or 2424298.1387$\pm$0.0084 for the $\dot{P}$=1.4$\times$10$^{-11}$ days per cycle case.  These two cases give the period in 1925.402 as 0.14531361$\pm$0.00000019 or 0.14531423$\pm$0.00000019 d respectively.  As expected, the period from the DFT exactly equals the period from the $\dot{P}$=0 fit.  As we expect that the $\dot{P}$ value is the same before and after eruption, we see that $P_{pre}$=0.14531423$\pm$0.00000019 d.  The folded light curve for all the pre-eruption magnitudes is displayed in Fig. 5.
		
\begin{figure}
	\includegraphics[width=\columnwidth]{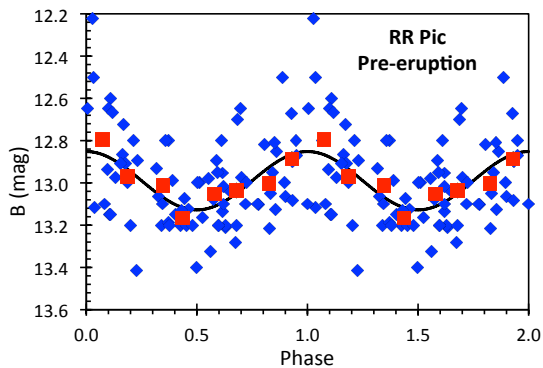}
    \caption{Folded light curve for all 82 pre-eruption magnitudes.  The individual magnitudes (blue diamonds) and the phased binned light curve (large red squares) are scattered around the best fit sinewave (black curve) with similar scatter as in Fig. 2 and 3.  The amplitude of this significant sinewave is 0.276 mag, closely similar to the average amplitude of the post-eruption light curve in quiescence.  This folded light curve is good looking, and with the expected amplitude and phase, hence giving good confidence that the periodicity is real.}
\end{figure}
	
	The epoch of maximum light at 1925.402 is better determined from the pre-eruption data (tucked up to 1925.135) than from the post-eruption data (extrapolated back from 1944).  From this, we get the point $O-C$=0.1608$\pm$0.0084 for 1925.402, and this is plotted in the post-eruption $O-C$ curve (Fig. 1).  Note, that it is fully possible to adopt a different cycle count, moving the point upwards or downwards by 0.145 days.  But such would then require RR Pic to have a large period change sometime before 1944.  So Occam's Razor points to the $O-C$ point as shown.  With this, we see a point modestly near to the parabolic and linear extrapolations.
	
	I have also fitted the first 41 magnitudes and the last 41 magnitudes.  Both of them show a good fit at near the overall period.  This provides confidence that the periodicity is real, and not just arising from some artifact.  This also provides two epochs for the middle of the two halves, as presented in Table 1. 

\subsection{RR Pic Joint Fit}

	The best $\Delta$P and $\dot{P}$ values will come from using all the data in a joint chi-square fit.  This is with one ephemeris, a broken parabola, for 82 pre-eruption magnitudes and 210 post-eruption times of maximum brightness.  The model has four fitting parameters ($P_{pre}$, $E_0$, $P_{post}$, and $\dot{P}$), so the fit has 288 degrees of freedom.  The best fit has a very narrow minimum in chi-square, at a value of 303.0.  The best fit has the $E_0$ value in the middle of the values derived for the pre- and post-eruption data.  The one-sigma error region is for sets of parameters that have the chi-square $<$304.0.
	
	The best fit has $P_{pre}$=0.14531434$\pm$0.00000013 and $P_{post}$=0.1450237620$\pm$0.0000000031.  The observed steady period change is $\dot{P}$=(+1.25$\pm$0.01)$\times$$10^{-11}$ days/cycle, while the epoch of maximum light at the time of the eruption is $E_0$=2424298.1600$\pm$0.0006.  The period change is $\Delta$P=$-$0.00029058$\pm$0.00000013 days, with the fractional change in period being $\Delta$P/P=$-$2003.7$\pm$0.9 ppm.

\subsection{$\Delta$P for RR Pic}

	RR Pic has $\Delta$P/P=$-$2003.7$\pm$0.9 ppm.  The negative sign shows that the orbital period {\it decreased} across the 1925 eruption.  With the orbital period getting smaller, the orbital separation must have experienced a sudden shortening, which is to say that the system got tighter and the companion star's Roche lobe must have gotten smaller.
	
	How confident can we be that RR Pic suffered this large negative period change?  Let me offer four ways of evaluating the confidence:  
	
	(1) The pre-nova system really must have shown sinusoidal modulation on the orbital period with roughly a quarter-magnitude amplitude, and such should be easily visible in an 82 magnitude light curve from Harvard.  The DFT is optimal and exhaustive for sinewave period discovery.  From the statistics of looking at noise peaks over a very wide range of periods, we know that noise peaks never get above 6.6 in Fig. 4.  (The probability of noise peaks falls off exponentially with a scale of the average power, 1.0 in the vertical axis, of Fig. 4.)  Simulations prove that a quarter-magnitude amplitude must produce a DFT peak with power $>$9 in the figure.  So we know that the true $P_{pre}$ must appear as the one isolated peak >9, far above the noise peaks $<$6.6.  And that is exactly what we see in the DFT, just one isolated peak greatly above the noise, so that must confidently point to the true period.
	
	(2) The pre-nova periodicity has its shape, amplitude, period, and phase as required for a true orbital modulation, whereas it is very unlikely that some noise or artifact would produce a false-periodicity as close as observed.  That is, the folded curve is approximately a sinewave, as seen for the post-eruption light curve, whereas some artifacts would produce effects like gaps in the phase curve for aliases or the clustering in phase of a few discrepant points that happened together for some false-period.  The observed amplitude (0.276$\pm$0.058 mag) is the same as required, based on the post-nova behavior, whereas the highest noise peak in the DFT has a substantially smaller amplitude.  If the periodicity were noise or the result of some artifact, then it would be likely to have a substantially different amplitude.  The orbital period is within 3000 ppm of $P_{post}$, as required to have some known physical mechanism produce such a change.  The phase of maximum light at the time of the nova ($O-C$=0.1608) deviates from the best extrapolation of the post-eruption data by 0.04 days, or about a quarter of a period, all within the error bars of the extrapolation.  This coincidence in phase points weakly to the periodicity as being real.  These four properties of the best fit match up with requirements for a real periodicity, whereas one or more of these properties would likely be off if the periodicity is not real.
	
	(3) We can perform an F-test comparison of the two models where we have a freely fit sinewave versus where the amplitude of that sinewave is zero.  By allowing for the one parameter to change, the chi-square goes from 104.3 for the zero amplitude case to 81.3  for the best fit case.  The chi-square improves by 23.0 for changing one parameter.  If the periodicity is not real, then the probability is 7$\times$10$^{-6}$ that any one period will produce such a large improvement in chi-square by changing only one model parameter.  For the observed width of the DFT peak, we have searched roughly 480 independent trial periods in our search.  This makes the probability of a false alarm as 0.0033.  For Gaussian statistics, this corresponds to a 3-sigma probability that the periodicity is not a false alarm.  This is not a very low probability (due to the large range for searching for trial periods), but it does satisfy the 3-sigma standard for being a significant result.
	
	(4) We can perform another F-test, comparing the best fit model versus the same model except where the period is allowed to vary over the search range.  The average chi-square for periods inside our search range is 128.  That is, the chi-square improves by 46.7 (from 128 to 81.3) by changing one parameter.  Such an improvement is very unlikely (at the 1.3$\times$10$^{-9}$ probability level) for random noise for a given trial period.  I examined roughly 480 independent trial periods, so the overall probability of non-periodic data producing such a large improvement in the chi-square is 6.5$\times$10$^{-7}$, which corresponds to a Gaussian significance of 5-sigma.
	
	In all, the pre-eruption periodicity is significant.  And the total uncertainties in $P_{pre}$ and $P_{post}$ are greatly smaller than their difference.  So we are left with a strong conclusion that $\Delta$P/P=$-$2003.7$\pm$0.9 ppm, with the value large and negative.

\section{HR Del}

HR Del erupted on 1967.425, reaching a peak of V=3.6, the fourth brightest classical nova since World War II (Strope et al. 2010).  It's light curve showed many short sporadic flares, or jitters, still of unknown origin, and hence has a J-class light curve.  HR Del has a {\it Gaia} DR2  distance of 958 parsecs (Schaefer 2018), and its quiescent counterpart is the second brightest of all classical novae (after V603 Aql) at V=12.1.  After the eruption, the quiescent counterpart is seen to have a sinusoidal brightness change with a period of 0.214164 days and a full amplitude of just under 0.1 mag (Friedjung, Dennefeld \& Voloshina 2010).

The high brightness of the quiescent nova means that we can get many archival plates from before the eruption.  With many such plates, a sinusoidal oscillation of 0.1 mag should be easy to pull out.  As such, HR Del becomes one of the few classical novae for which I can seek to measure $\Delta$P.

\subsection{HR Del Post-Eruption Orbital Period}

The post-eruption orbital period has been measured a number of times:  Bruch (1982) made a radial velocity curve with $P_{post}$=0.2141674$\pm$0.0000002 days.  K$\ddot{\rm u}$rster \& Barwig (1988) also present a radial velocity curve with $P_{post}$=0.214165$\pm$0.000005 days.  Friedjung et al. (2010) report $P_{post}$=0.214164 days.  McQuillen et al. (2012) gives $P_{post}$=0.21423$\pm$0.00005 days the sinusoidal photometric modulation.  These periods are consistent and they are adequately accurate for the purposes of this paper.

Nevertheless, it is good to get a more accurate period, especially so as to seek any period changes, so as to derive the best period just after the eruption.  The only way to improve the $P_{post}$ measure for extrapolation back to 1967.425 is to collect many epochs of maximum photometric brightness and plot the $O-C$ curve.  For this, I have observed the light curve in 2013 at the Highland Road Park Observatory (HRPO), analyzed the AAVSO light curves for 2014, 2016, and 2017 plus the SuperWASP light curves for 2004 and 2006, as well as the ASAS light curve from 2003--2009.  These epochs and the measured periods are listed in HJD in Table 3.

\begin{table*}
	\centering
	\caption{Periods and Times of Maximum Light for HR Del}
	\begin{tabular}{lllll} 
		\hline
		Data Source	&   Year	&   $P_{post}$ (days)   & $T_{max}$ (HJD) & $O-C$ (days)  \\
		\hline
Radial velocity data	&	1979.71	&	0.2141647	$\pm$	0.0000008	&	...			&	...			\\
SuperWASP (2004, 2006)	&	2005.48	&	0.21416290	$\pm$	0.00000029	&	...			&	...			\\
SuperWASP (2004)	&	2004.58	&	...			&	2453219.0442	$\pm$	0.0007	&	0.0039	$\pm$	0.0050	\\
SuperWASP (2006)	&	2006.65	&	...			&	2453974.1825	$\pm$	0.0007	&	-0.0039	$\pm$	0.0050	\\
ASAS (2003--2009)	&	2006.54	&	0.2141650	$\pm$	0.0000017	&	2453933.4920	$\pm$	0.0054	&	-0.0030	$\pm$	0.0074	\\
HRPO (2013)	&	2013.87	&	...			&	2456610.5521	$\pm$	0.0010	&	-0.0064	$\pm$	0.0051	\\
AAVSO (2014--2017)	&	2016.88	&	0.21416372	$\pm$	0.00000030	&	...			&	...			\\
AAVSO (2014)	&	2014.58	&	...			&	2456869.0627	$\pm$	0.0008	&	0.0069	$\pm$	0.0051	\\
AAVSO (2016)	&	2016.61	&	...			&	2457611.1395	$\pm$	0.0008	&	0.0017	$\pm$	0.0051	\\
AAVSO (2017)	&	2017.62	&	...			&	2457979.0726	$\pm$	0.0005	&	-0.0008	$\pm$	0.0050	\\
		\hline
	\end{tabular}
\end{table*}

My observations at HRPO (in suburban Baton Rouge Louisiana) were made with the 20-inch telescope with a V filter.  The extraction of the magnitudes was with aperture photometry using IRAF and nearby comparison stars with magnitudes from APASS.  Time series were made on 10 nights over a 42 day interval in October and November of 2013, for a total of 1261 magnitudes.  This light curve was fit to a sinusoid with a chi-square method.  The full amplitude was 0.124 mag in the V-band.  The folded light curve and this best fit is shown in Fig. 6.  The time interval is too short to get a period of useful accuracy, but this does give an excellent epoch (see Table 3). 
		
\begin{figure}
	\includegraphics[width=\columnwidth]{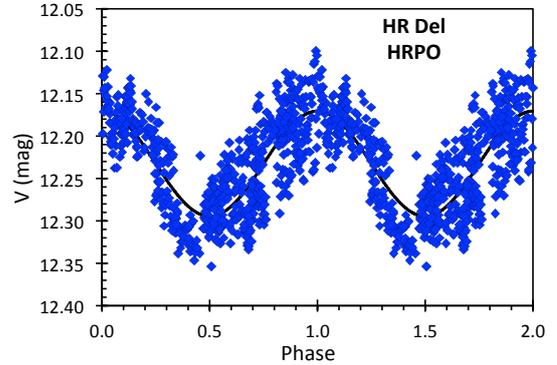}
    \caption{Folded light curve for all 1261 post-eruption V magnitudes from HRPO.  We see the typical sinusoidal oscillations (0.124 mag peak-to-peak) with large amplitude flickering.  The best fit sinewave is shown by the black curve, and the epoch of maximum brightness is taken from the fit for a time near the middle of the October--November 2013 observing run.}
\end{figure}

The AAVSO database contains many excellent time series, with these yielding accurate periods and epochs.  I have used only the V-band CCD observations, with these clustered almost exclusively in the years 2014, 2016, and 2017.  Many observers contributed 2176, 3728, and 8939 magnitudes in the three years respectively.  The average folded light curves appear as close to a sinewave, and I made chi-square fits.  The best fit period for all the data 2014--2017 is 0.21416372$\pm$0.00000030 days.  The three best fit epochs for each year are presented in Table 3.

SuperWASP (Pollacco et al. 2006) observed 3896 unfiltered CCD magnitudes in 2004 and 2006 with an array of 200 mm telephoto lenses placed at the Roque de los Muchachos observatory in the Canary Islands.  The SuperWASP light curve has already been presented in McQuillin et al. (2012), and their Fig. 3 shows the folded light curve, with the usual large scatter (from flickering) superposed on an apparently-perfect sinewave with amplitude $\sim$0.1 mag.  I have fit sine waves to the 8925 SuperWASP magnitudes, with the best fit period being 0.21416290$\pm$0.00000029 days for the combined 2004 and 2006 data sets.  This is very close, but about 2-sigma different, to the AAVSO period.  The best fit epochs of maximum brightness in the fitted sine wave are tabulated in Table 3.

ASAS reports 300 V magnitudes from 2003--2009, all taken with a 200 mm f/2.8 lens at Las Campanas Observatory in Chile.  My best fit sinewave has a full amplitude of 0.096 mag$\pm$0.016 mag and a period of 0.2141650$\pm$0.0000017 days.  The epoch for the maximum brightness of the best fit sinewave close to the middle of the observing interval is given in Table 3.

A variety of data sources are not useful for the purposes of constructing an $O-C$ curve:  (1) Kohoutek \& Pauls (1980) and Kohoutek, Pauls \& Steinbach (1981) present times of a particular photometric phase from 1978--1980, and this would hopefully be a wonderful way to pin down the early $O-C$ curve.  But their times of maximum and minimum show variations by over 0.25 in phase, for any plausible period.  A look at their light curves (Kohoutek et al. 1981), shows that their epochs are based on single cycles, and such is seen to have huge real uncertainties in defining any phase.  And they expressed their epochs only for a set of alias periods, with the quoted epochs changing greatly for each adopted incorrect period.  Further, their three quoted epochs for each year are inconsistent with each other up to 0.41 in phase.  All this is to say that we cannot use their epochs.  (2) The epoch in Fig. 6 of Friedjung et al. (2010) is ambiguous and inconsistent with their Fig. 2, so this cannot be used.  (3) The AAVSO visual light curve cannot be used because their timing and photometric accuracy is too poor.  (4) There is no significant periodicity in the light curves from Barnes \& Evans (1970), Dreschel \& Rahe (1980), and the Harvard Damon plates late in the tail of the eruption, variously because the eruption light dominates or the data set has too few observations for their photometric accuracy.  (5)  Bruch (1982) and K$\ddot{\rm u}$rster \& Barwig (1988) both quote epochs of maximum radial velocity, but the offset in phase to get to the epochs of maximum brightness is poorly known.  In an idealized binary system, the offsets would be accurately predicted, but HR Del has hot spots in the disk making for different offsets for both the radial velocity curve and for the photometric modulation.  K$\ddot{\rm u}$rster \& Barwig use some unstated light curves to claim that the maximum brightness is at phase 0.7$\pm$0.1 (and my best fit ephemeris, see below, suggests a phase of 0.6), but the error bars are too big on this to be useful, and adopting some value for the offset would now be tantamount to assuming that which we seek to measure.  So the radial velocity data are not useful for the purpose of measuring epochs for the $O-C$ curve.  

The radial velocity data cannot yield a useable epoch for the $O-C$ curve, but it can provide a nice period.  I have collected a total of 120 radial velocity measures of the emission line for He {\rm II} from 1977--1980 from Bruch (1982), K$\ddot{\rm u}$rster \& Barwig (1988), and Hutchings (1979; 1980).  These form a well-measured sine wave.  With the amplitude of the radial velocity curve ($K$=109$\pm$3 km s$^{-1}$) being large compared to the scatter, we can get a good period just from the 1977--1980 data.  With a chi-square fit, I get a period of 0.2141647$\pm$0.0000008 days.

The usable epochs and periods are collected into Table 3, and these are made into an $O-C$ curve in Fig. 7.   The key point is that we have no epochs before 2004, and it is a long extrapolation back to 1967.425.  So we already know that we cannot confidently determine an epoch for the time of eruption (for use with measuring $P_{pre}$).  Further, the $O-C$ curve appears to jerk up and down with an amplitude much larger than the measurement errors.  So, apparently, HR Del is like DQ Her in that it suffers apparent period changes on a time scale faster than a year, with these jerks in the $O-C$ curve being to both longer and shorter periods.  Critically, the total amplitude of these $O-C$ excursions is actually quite small, appearing no more than 0.01 days in size.
		
\begin{figure}
	\includegraphics[width=\columnwidth]{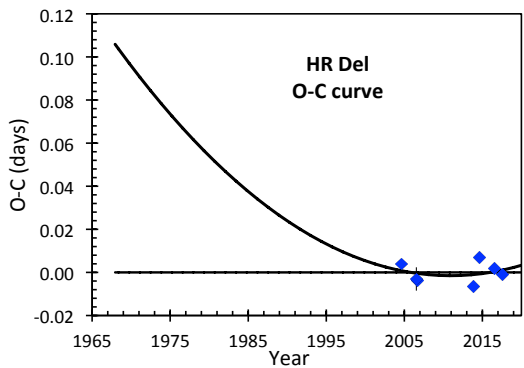}
    \caption{HR Del post-eruption $O-C$ curve.  The seven useable epochs of maximum brightness are plotted as the deviation of these times from a fiducial linear ephemeris.  The best fit linear ephemeris has a period of 0.21416508 days and an epoch of HJD 2453974.1864, and this ephemeris is used as the fiducial model for constructing the $O-C$ curve.  As such, this linear model is a flat line at $O-C$=0 extrapolated back to the nova eruption in 1967.425.  The variations of the measured points about this best fit line is much larger than the likely real error, showing that the period suffers very small changes on time scales of under a year.  Unfortunately, the post-eruption data only extend from 2004--2017, so a substantial parabolic term is possible.  The post-eruption curve must link up continuously with the pre-eruption curve.  The best fit pre-eruption data have an epoch in 1967.425 that is about half an orbital period away from the flat model, so we can exclude the zero-$\dot{P}$ case.  The pre-eruption light curve is fitted to prefer a positive-$\dot{P}$.  The joint best fit for both the pre- and post-eruption data result in $\dot{P}$=(4.0$\pm$0.9)$\times$10$^{-11}$ days/cycle.  The parabolic curve displays the post-eruption portion of this joint fit.}
\end{figure}

We have three highly accurate measures of the orbital period from the data sets that span a few years; for the radial velocity curve 1977--1980, the SuperWASP photometry 2004--2006, and the AAVSO light curves from 2014--2017.  These periods do not progress in any steady manner such as expected for a steady $\dot{P}$.  The periods differ from each other at roughly the 2-sigma level.  With this and the placement in the $O-C$ curve, there must have been at least two very small period changes between 2006 and 2014.  So both the periods and the epochs are pointing to very small period changes on times of a few years and shorter.

I have made chi-square fits to the $O-C$ curve from 2004--2017.  For this, I have added in quadrature a systematic error of 0.005 days to reflect the reality that the individual measures have a scatter (either from systematic error or from intrinsic variations) past what any smooth curve can provide.  The best fit linear ephemeris has $P_{post}$=0.21416508$\pm$0.00000016 d and an epoch of HJD 2453974.1864$\pm$0.0020.  This best fit is taken as the model for constructing the $O-C$ curve in Fig. 7.  This best fit can be extrapolated back to form an epoch close to the time of the eruption (1967.425) as HJD 2439838.0061$\pm$0.0107.

Does HR Del have a significant steady period change (i.e., $\dot{P}$) in the post eruption time interval?  This is hard to say.  The main problem is that I only have the $O-C$ curve from 2004--2017.  Over this time interval, a fitted parabola implies that |$\dot{P}$|$\lesssim$10$^{-10}$ days/cycle.  The measured periods from 2004--2006 and 2014--2017 differ by just 4 ppm.  The $P$ from the radial velocity curves for 1977-1980 has too large an uncertainty to usefully constrain the $\dot{P}$ term.

How can we extrapolate the $O-C$ curve back to 1967.425?  There are many reasonable possibilities, with two included as curves in Fig. 7.  (1) A simple best fit linear model extrapolates back to HJD 2439838.0061 with $P_{post}$=0.21416508$\pm$0.00000016 d.  But this solution will be inconsistent with the pre-eruption fit.  (2)  We can also come up with steady-$\dot{P}$ models (i.e., parabolas in the $O-C$ diagram) that match up with the epoch returned by the pre-eruption analysis.  From the joint fit in Section 3.4, I will find that the best epoch at the time of the eruption for a minimal positive $\dot{P}$=-4.0$\times$10$^{-11}$ days/cycle is HJD 2439838.1120.  With the post-eruption parabola passing through this point, the value for $P_{post}$ is 0.21416215$\pm$0.00000055 d.  (3) We can also get a solution for the joint fit when the curvature is negative, where the post-eruption parabola passes 1967.425 roughly one period before the best fit case.  That is, the cycle count from 1967.425 to 2004 differs by one from the positive-$\dot{P}$ solution.  But the pre-eruption light curve is relatively poor for this required negative-$\dot{P}$, so we can reject this possibility.

With the post-eruption data alone, we have no effective extrapolation back to the year of the nova (because we do not have any useful constraint on $\dot{P}$), but we do have a sure knowledge of $P_{post}$ to within 15 ppm or so.  The value of $\dot{P}$ will be constrained fairly tightly with the 66 years of pre-eruption light curve, and then pinned down when the pre- and post-eruption $O-C$ curves are required to meet in 1967.

\subsection{HR Del Pre-Eruption Magnitudes}

The only way to get pre-eruption magnitudes, so as to measure $P_{pre}$, is to use the archival plates at Harvard and Sonneberg.  To this end, I have visited Harvard three times and Sonneberg twice, independently measuring the B magnitudes of HR Del on each visit.  On these various visits, only about half of the available plates were measured during a visit, so each individual plate has from one-to-three independent measures.  The magnitudes were measured visually from each plate, with APASS comparison stars, and then averaged together for measures of each plate.  

Schweitzer (1968) reports on eight pre-nova magnitudes from 1963--1965 as taken from blue-sensitive archival plates taken with the 0.5-m telescope at Strasbourg Observatory.  The adopted magnitude of their comparison stars are given, and these are 0.06--0.08 mag too-bright in the relevant range.  With the modern magnitudes for these stars from APASS and the conversion equation given in Johnson, et al. (2014), these magnitudes were converted to modern B magnitudes.  Two of the reported magnitudes are far outliers, at 11.51 and 11.76, with such brightness being either some measurement error or some rare intrinsic variations of HR Del that have nothing to do with the orbital modulation, so these two magnitudes have been disregarded.

The result is 495 pre-nova B-band magnitudes; including 135 from Harvard from 1901--1953, 354 from Sonneberg from 1940 to May 1967, plus 6 from Schweitzer from 1963--1965.  These are listed in Table 4, with the full table only appearing in the electronic version of this paper.

\begin{table}
	\centering
	\caption{HR Del pre-eruption B magnitudes (full table with 495 magnitudes is on-line only)}
	\begin{tabular}{llll} 
		\hline
		Mid-exposure (HJD)	&   Year   & B (mag) & Source  \\
		\hline
2415693.4707	&	1901.844	&	11.81	&	HCO (AC 1954)	\\
2415723.5179	&	1901.926	&	12.13	&	HCO (AC 2043)	\\
2416033.5585	&	1902.776	&	12.39	&	HCO (AC 2888)	\\
2416061.5155	&	1902.852	&	12.01	&	HCO (AC 2978)	\\
2416183.8956	&	1903.187	&	12.26	&	HCO (I 30245)	\\
...	&	...	&	...	&	...	\\
2439436.2894	&	1966.848	&	12.13	&	Sonneberg	\\
2439443.3028	&	1966.867	&	12.42	&	Sonneberg	\\
2439596.5951	&	1967.287	&	12.33	&	Sonneberg	\\
2439618.5418	&	1967.347	&	12.34	&	Sonneberg	\\
2439621.5401	&	1967.355	&	12.45	&	Sonneberg	\\
		\hline
	\end{tabular}
\end{table}

The average scatter of these measures is around 0.2 mag, after allowing for shifting average levels, with this being somewhat larger than expected for such a bright star.  The magnitudes have some apparent quantization because one of the visits had the magnitude of HR Del estimated only to the nearest tenth-magnitude.  The Sonneberg plates show an apparent small and fast jump by a quarter of a magnitude around 1956.  Such jumps are not seen for other CV light curves that I or others have made at Sonneberg, so it is likely not instrumental.  The default idea would be that HR Del actually suffered a small state change around 1956, with such not being testable in the Harvard data due to the Menzel Gap.  The scatter in the best folded light curves is substantially larger than the expected amplitude of close to 0.10 mag, so the folded light curves look poor, with the sinusoidal modulation not readily apparent by eye.  However, the big advantage for HR Del is that it is bright, so I can get many pre-eruption plates and that makes for the orbital period to be detectable with the usual statistical tools.  

Within a year or two, the DASCH program will release their photometry of $>$1000 plates.  HR Del is a perfect target for DASCH, where it is exhaustive at pulling out all plates, the DASCH program will yield magnitudes for large numbers of plates, many more than I have visually  measured.  Further, as a bright target with no crowding, DASCH will perform excellent photometry.  So we can anticipate a substantial improvement in the pre-eruption light curve within a few years.

\subsection{HR Del Pre-Eruption Orbital Period}

The sinusoidal photometric modulation has a small amplitude, close to 0.10 mag, embedded in a noisy light curve, so the period can only be plucked out by statistical methods.  

\subsubsection{Fourier Transform}

The optimal method for finding a sinewave periodicity in sparsely sampled data is to use a discrete Fourier transform (DFT).  (This is true for the $\dot{P}$=0 case found in the previous section.  And it is still true for the low-$\dot{P}$ cases of relevance here.)  The physically plausible range for $P_{pre}$ is within 3000 ppm of $P_{post}$, or from 0.2135 to 0.2148.  I heavily oversampled the DFT, with 13,000 trial periods in the search range.  The DFT for HR Del is shown in Fig. 8.
		
\begin{figure}
	\includegraphics[width=\columnwidth]{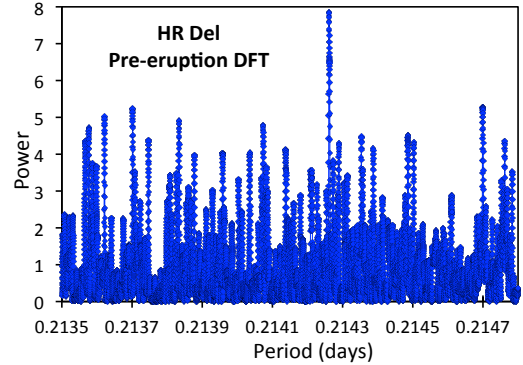}
    \caption{HR Del pre-eruption DFT.  This period search from 0.2135--0.2148 days shows just one high peak that is blatantly far above all background noise peaks.  So we know that $P_{pre}$=0.21426141 days.}
\end{figure}

The DFT shows only one peak greatly higher than the highest noise peaks.  (Indeed, this peak is the highest peak over a very wide range of trial periods, with the second highest peak simply being the monthly alias of the highest peak.  The yearly alias also has a peak with a power comparable to the taller noise peaks.)  The period is 0.21426141 days.  This peak is far above the background noise peaks.  That is, the DFT power is 7.9 for the high peak, and 5.2 for the next highest peak, while the average power is 0.93.

The FWHM of the peak is 0.0000040 days.  The period search range is 0.0013 days.  So there are 325 independent trial periods examined.

The DFT shows a high peak far above the noise peaks with a period in the search range, corresponding to a sinewave with a full amplitude of 0.12 mag.  This is exactly as expected.  That is, the post-eruption quiescence shows a sinewave with an amplitude that is typically 0.10 mag, so it is inevitable that the pre-eruption quiescence must also show a sinewave with an amplitude near 0.10 mag and within the period search range.  Indeed, if the DFT peak is not the true $P_{pre}$, then we have to ask where is the required peak in the DFT for the true period?  That is, the true $P_{pre}$ must produce a high peak in the DFT, and there is only one such peak, so that peak must be the true $P_{pre}$.

\subsubsection{Chi-Square Fits}

The DFT does not readily produce the epoch, amplitude, the error bars, or the significance of the periodicity.  But a chi-square fit solves all these problems.  So I have fitted a sinewave to the 495 pre-eruption magnitudes with a chi-square analysis.  Two differences with the prior cases (DQ Her, BT Mon, and RR Pic) is that I cannot usefully give the epoch at the time of the eruption, and I have little useful information on $\dot{P}$ from the post-eruption data alone.  The sigma in the denominator of the chi-square arises from both the measurement errors and the flickering, so I can only set it to the RMS scatter about the best fit curve, or 0.23 mag.

For the case with an assumed $\dot{P}$=0, the chi-square has a blatant minimum for a period of 0.21426137$\pm$0.00000019 d.  The epoch for this linear ephemeris is HJD 2433720.0674$\pm$0.0089, which translates to the date of the eclipse to $E_0$=2439838.0866$\pm$0.0104.  The full amplitude is 0.12$\pm$0.03 mag.  The folded light curve with all 495 points does not readily show the periodicity due to the large scatter, but when the points are binned in phase, we see a good sinewave.  The chi-square is 510.5 (with 495-2 degrees of freedom), with this by construction as the uncertainty is set to 0.23 mag.

If we allow $\dot{P}$ to be a free parameter, we can get a slightly better fit.  The best fit has a chi-square of 507.0 for $\dot{P}$=+3$\times$10$^{-11}$ days/cycle.  This solution has $P_{pre}$=0.21426267 d and $E_0$=2439838.108$\pm$0.012.  The formal 1-sigma range (with chi-square values within 1.0 of the minimum) is from (+1 to +6)$\times$10$^{-11}$ days/cycle, while the 2-sigma range (with chi-square values within 4.0 of the minimum) is from (-2 to +7)$\times$10$^{-11}$ days/cycle.  

We can test the periodicity for being a false alarm with an F-test comparing a model with a variable amplitude versus a model with the amplitude set equal to zero.  That is, how does the best fit model for the pre-eruption light curve with something like the expected post-eruption amplitude compare with a model where there is no periodicity?  For the addition of one free fit parameter (the amplitude is allowed to vary), the chi-square improves by 20.7 (with 492 degrees of freedom).  This is a large improvement for adding just one parameter, with this being very unlikely to happen for data that do not have a true periodicity.  The F-test probability is 1-6.7$\times$10$^{-6}$.  To get this highly improbable improvement in chi-square (highly improbable for a false alarm), I examined 325 independent trial periods.  This makes for an improbability of 0.00217.  This corresponds to the Gaussian probability of 3.1-sigma.  That is, the existence of the periodicity is at the 3.1-sigma level.  While this is not at a very high confidence level, it still passes the traditional 3-sigma requirement, so I can conclude that this per-eruption periodicity is significant.

We can further make an F-Test comparing the best fit model versus the same model except that the period is allowed to vary.  This is asking a subtly different question than the previous F-Test, but it is still essentially asking for the significance of the existence of the pre-eruption periodicity.  The average chi-square over the period search interval is 549.  So a random trial period for which there is no real periodicity is worse than the best fit model by a chi-square difference of 42, with 492 degrees of freedom.  Such an improvement is very unlikely unless that best fit model has a true underlying periodicity.  The improbability of a trial period producing such a good improvement is 2.2$\times$10$^{-10}$.  To get this result, I have examined 325 independent test periods.  So the improbability that any of these test periods will produce such a large improvement in chi-square (in the absence of a true periodicity being present) is 7.2$\times$10$^{-8}$.  This corresponds to a Gaussian 5.5-sigma confidence level.  That is, the pre-eruption periodicity is significant at the 5.5-sigma level, and can easily be called `confident'.

\subsection{HR Del Joint Fit}

The $O-C$ curve for HR Del must be continuous across the eruption, which is to say that the stars do not jump forwards or backwards in their orbits.  (The period can change fast, but this comes out in the $O-C$ diagram as a sudden change of slope.)  This forces a connection between the pre-eruption fit and the post-eruption fit, where the two fits must meet at the date of the eruption.  That is, $E_0$ for the before and after fits must be equal.  So we need a joint fit involving both the 495 pre-eruption magnitudes and the 7 post-eruption epochs of maximum brightness.

For the model, we have only poor constraints on $\dot{P}$ from either the pre- and post-eruption data.  If both are allowed to be freely changing, then we could find reasonable solutions for most any value of $E_0$.  But the physics of any steady period change before the eruption is likely to be essentially equal to the physics of the steady period change after the eruption.  That is, $\dot{P}$ is likely to be a single constant both before and after the eruption.  In my joint fits, I will adopt this more likely and more restrictive assumption.

The joint model then has just four parameters; $P_{pre}$, $E_0$, $P_{post}$, and $\dot{P}$.  (The sinewave amplitude is held constant, and it does not change significantly over the ranges of interest.)  The data points number 495+7, and there are 4 fit parameters, so my joint fit has 498 degrees of freedom.

The key difference between the joint fit and the pre- and post-eruption fits is the requirement that both share the identical $E_0$.  If $\dot{P}$=0, then the before and after fits are a mismatch by about half an orbital phase, and this is much larger than the uncertainty in the two extrapolated values of $E_0$.  The pre-eruption light curve is pointing to $\dot{P}$$\sim$3$\times$10$^{-11}$ days/cycle (i.e., curvature in the $O-C$ diagram with the concave side up).  This produces a value for $E_0$ near HJD 2439838.12.   The post-eruption $O-C$ curve also runs through this epoch and curves down to be nearly flat in modern times (see Fig. 7), thus this appears to be a good solution.  For the post-eruption cycle count, it might be that the $E_0$ value is one orbital period earlier, but this would then require a negative-$\dot{P}$ for the post-eruption fit, and such a negative value is inconsistent with the pre-eruption constraint.  So, by combining the pre- and post-eruption data sets, we get a unique value for $\dot{P}$ and a unique overall solution.  

The best joint fit returns orbital periods just before and after the nova event, with $P_{pre}$=0.21426326$\pm$0.00000055 d and $P_{post}$=0.21416215$\pm$0.00000048 d.  The value of $\Delta$P/P is $-$472.1 ppm, with the 1-sigma error bar equal to 4.8 ppm, somewhat larger than for simple propagation of errors.  The best fit $\dot{P}$ is (+4.0$\pm$0.9)$\times$10$^{-11}$ days/cycle.  The best fit joint epoch is $E_0$ at HJD 2439838.1120$\pm$0.0150, and the best fit full amplitude is 0.134$\pm$0.030 mag.  The minimum chi-square is 511.2, with 506.6 from the 495 pre-eruption magnitudes and 4.6 from the 7 post-eruption epochs of maximum brightness.  The resultant folded and binned light curve shows a good sinewave shape.  

\subsection{$\Delta$P for HR Del}

So we have measured $P_{pre}$ and $P_{post}$ finding that $\Delta$P/P is $-$472.1$\pm$4.8 ppm.  That is, the orbital period of HR Del {\it decreased} across its 1967 eruption.

How confident should we be in this $\Delta$P/P?  Well, the real uncertainties in $P_{post}$ and $\dot{P}$ are all greatly too small to matter, the formal uncertainty in $P_{pre}$ is also greatly too small to matter, so the only way to question the $\Delta$P/P value is to wonder whether the pre-eruption periodicity is true.  After all, the folded light curve does have large scatter (RMS of 0.23 mag) so that the sinewave is not readily visible to the eye.  But we have 495 pre-eruption magnitudes, and the phase-binned folded light curve does show a nice sinewave.  In all, there are four good reasons to know that the periodicity is true:  (1 and 2) The F-tests return significances at the 3.1-sigma and 5.5-sigma confidence levels, even after accounting for all the trial periods.  This is adequate to be confident that the periodicity is significant.  (3) The DFT shows a single peak in the physically plausible search range, far above the highest noise peaks.  A glance at this DFT by any worker with an experienced eye shows that the period is obviously the true period.  (4) We strongly expect that the pre-eruption quiescence must have the same sinewave modulation with an amplitude of around 0.10 mag as is always seen in the post-eruption quiescence, and that period must be within something like 3000 ppm of $P_{post}$.  And that is exactly what we see for HR Del before 1967.  So if the 0.21426326 d periodicity is not the true underlying period, then there really must be some other comparable DFT peak in addition.  But there is only one such peak.  That is, if the 0.21426326 d peak is not from the orbital modulation, then we have no way of understanding why there isn't a second similar peak.  Or, with only one peak appropriate for the known behavior in quiescence, that peak must be the true periodicity caused by the orbital modulation.  For any or all of these four reasons, we can be confident that the pre-eruption periodicity is the correct and true measure of the orbital period, and further that the $\Delta$P/P value is correct to within the stated error bars.

\section{Overview}

My $\Delta$P program now has measures of $\Delta$P and $\dot{P}$ for six CNe.  These results are collected in the first block of Table 5.

Salazar et al. (2017) reports a formal value of $\dot{P}$=(+6.4$\pm$2.4)$\times$10$^{-8}$ days/cycle for V1017 Sgr.  (The units are days/cycle, not days/day as stated one place in the table header.)  This is a huge value, rather hard to understand in its size.  However, the measured error bar is also huge, being comparable in size to the quoted value itself.  (The chi-square only improved by 1.55 for adding the extra $\dot{P}$ term.)  So the real $\dot{P}$ could well be small (comparable to the various model values and to the observed values for the other CNe), or even negative with a large size.  This is just saying that the V1017 Sgr data do not place a useful limit on $\dot{P}$, so I have not recorded this value in Table 5.

\begin{table*}
	\centering
	\caption{Overview}
	\begin{tabular}{lllllll} 
		\hline
		& QZ Aur	&   HR Del	&   DQ Her   & BT Mon & RR Pic	&   V1017 Sgr  \\
		\hline
\underline{{\bf $\Delta$P Program Results:}}	&				&				&				&				&				&				\\
~~Pre-eruption plates	&	60 (4 eclipses)			&	495			&	52 (3 eclipses)			&	90 (10 eclipses)			&	82			&	15			\\
~~$P_{pre}$ (d)	&	0.35760096			&	0.21426326			&	0.1936217610			&	0.33380167			&	0.14531434			&	5.787616			\\
	&		$\pm$	0.00000005	&		$\pm$	0.00000055	&		$\pm$	0.0000000055	&		$\pm$	0.00000011	&		$\pm$	0.00000013	&		$\pm$	0.000272	\\
~~$P_{post}$ (d)	&	0.35749703			&	0.21416215			&	0.1936208977			&	0.33381490			&	0.1450237620			&	5.786038			\\
	&		$\pm$	0.00000005	&		$\pm$	0.00000048	&		$\pm$	0.0000000017	&		$\pm$	0.00000006	&		$\pm$	0.0000000031	&		$\pm$	0.000078	\\
~~$\Delta$P (d)	&	-0.00010393			&	-0.00010111			&	-0.0000008633			&	0.00001323			&	-0.00029058			&	-0.001578			\\
	&		$\pm$	0.00000007	&		$\pm$	0.00000073	&		$\pm$	0.0000000058	&		$\pm$	0.00000013	&		$\pm$	0.00000013	&		$\pm$	0.000283	\\
~~$\Delta$P/P (ppm)	&	-290.71	$\pm$	0.28	&	-472.1	$\pm$	4.8	&	-4.46	$\pm$	0.03	&	$+$39.6	$\pm$	0.5	&	-2003.7	$\pm$	0.9	&	-273	$\pm$	61	\\
~~$\dot{P}$ ($10^{-11}$ days/cycle)	&	-2.84	$\pm$	0.22	&	$+$4.0	$\pm$	0.9	&	0.00	$\pm$	0.02	&	-2.3	$\pm$	0.1	&	$+$1.25	$\pm$	0.01	&	...			\\
\underline{{\bf System Parameters:}}	&				&				&				&				&				&				\\
~~Light curve class	&	S(25)			&	J(231)			&	D(100)			&	F(182)			&	J(122)			&	S(120)			\\
~~Magnetic WD?	&	Non-magnetic			&	Non-magnetic			&	Magnetic			&	Magnetic			&	Magnetic			&	Non-magnetic			\\
~~Distance (pc)	&	3200		$^{+4030}_{-330}$	&	958		$^{+35}_{-29}$	&	501	$\pm$	6	&	1477		$^{+128}_{-84}$	&	511	$\pm$	8	&	1269		$^{+84}_{-60}$	\\
~~$A_V$ (mag)	&	1.7			&	0.5			&	0.2			&	0.7			&	0.0			&	1.2			\\
~~$V_q$ (mag)	&	17.0			&	12.1			&	14.3			&	15.7			&	12.2			&	13.5			\\
~~$M_{V,q}$ (mag)	&	2.7			&	1.7			&	5.6			&	4.1			&	3.7			&	1.8			\\
~~$\dot{M}$ (M$_{\odot}$/year)	&	3$\times$10$^{-8}$			&	8$\times$10$^{-8}$			&	2$\times$10$^{-9}$			&	9$\times$10$^{-9}$			&	1$\times$10$^{-7}$			&	8$\times$10$^{-8}$			\\
~~$\dot{M}_{mb,model}$ (M$_{\odot}$/year)	&	2$\times$10$^{-8}$			&	3$\times$10$^{-9}$			&	2$\times$10$^{-9}$			&	1$\times$10$^{-8}$			&	1$\times$10$^{-9}$			&	...			\\
~~$M_{WD}$ (M$_{\odot}$)	&	0.98	$\pm$	0.045	&	0.67	$\pm$	0.08	&	0.6	$\pm$	0.07	&	1.04	$\pm$	0.06	&	0.95			&	1.1			\\
~~$M_{comp}$ (M$_{\odot}$)	&	0.93			&	0.55	$\pm$	0.03	&	0.4	$\pm$	0.05	&	0.87	$\pm$	0.06	&	0.4			&	0.6			\\
~~$q$	&	0.95	$\pm$	0.05	&	0.82	$\pm$	0.11	&	0.67	$\pm$	0.11	&	0.84	$\pm$	0.08	&	0.4			&	0.5			\\
~~$a$ (R$_{\odot}$)	&	2.56			&	1.57			&	1.37			&	2.45			&	1.25			&	15.74			\\
~~$R_{comp}$ (R$_{\odot}$)	&	1.03	$\pm$	0.01	&	0.56			&	0.47			&	0.87			&	0.39			&	5.14			\\
~~$V_{comp}$ (km s$^{-1}$)	&	186			&	204			&	215			&	202			&	307			&	89			\\
~~$V_{WD}$ (km s$^{-1}$)	&	176			&	167			&	144			&	169			&	129			&	49			\\
~~$V_{ejecta}$ (km s$^{-1}$)	&	1000			&	525			&	800			&	2100			&	850			&	1000			\\
~~$M_{ejecta}$ (M$_{\odot}$)	&	1$\times$10$^{-5}$			&	1$\times$10$^{-4}$			&	1$\times$10$^{-4}$			&	2$\times$10$^{-5}$			&	2$\times$10$^{-5}$			&	6$\times$10$^{-6}$			\\
~~$M_{ignition}$ (M$_{\odot}$)	&	6$\times$10$^{-6}$			&	3$\times$10$^{-5}$			&	8$\times$10$^{-5}$			&	8$\times$10$^{-6}$			&	6$\times$10$^{-6}$			&	4$\times$10$^{-6}$			\\
~~$\tau_{rec}$ (yr)	&	300			&	400			&	40000			&	2000			&	200			&	100			\\
~~$T_{comp}$ (K)	&	5200	$\pm$	200	&	3700			&	3300			&	5300			&	3400			&	5200			\\
~~$H$ (km)	&	180			&	63			&	55			&	140			&	38			&	6900			\\
\underline{{\bf $\Delta$P calculations:}}	&				&				&				&				&				&				\\
~~$\Delta$P/P observed (ppm)	&	-290.71	$\pm$	0.28	&	-472.1	$\pm$	4.8	&	-4.46	$\pm$	0.03	&	$+$39.6	$\pm$	0.5	&	-2005	$\pm$	12	&	-273	$\pm$	61	\\
~~$\Delta$P$_{ml}$/P (ppm)	&	10			&	164			&	200			&	21			&	30			&	7			\\
~~$\Delta$P$_{FAML}$/P (ppm)	&	-0.2			&	-6.6			&	-5.9			&	-0.2			&	-1.3			&	-0.07			\\
~~Max. $\Delta$P$_{jet}$/P (ppm)	&	$\pm$42			&	$\pm$320			&	$\pm$560			&	$\pm$160			&	$\pm$61			&	$\pm$60			\\
~~$\Delta$P$_{5mag}$ (ppm)	&	1720			&	1030			&	1030			&	1580			&	960			&	13400			\\
~~$\Delta$P$_{M_q=+12}$/P (ppm)	&	3200			&	2230			&	1360			&	2500			&	1600			&	27300			\\
\underline{{\bf $\dot{P}$ calculations:}}	&				&				&				&				&				&				\\
~~$\dot{P}$ observed ($10^{-11}$ days/cycle)	&	-2.84	$\pm$	0.22	&	$+$4.0	$\pm$	0.9	&	0.00	$\pm$	0.02	&	-2.3	$\pm$	0.1	&	$+$1.25	$\pm$	0.01	&	... 			\\
~~$\dot{P}_{model}$ ($10^{-11}$ days/cycle)	&	-0.54			&	-0.043			&	-0.027			&	-0.33			&	-0.007			&	...			\\
~~$\dot{P}_{mt}$ ($10^{-11}$ days/cycle)	&	0.19			&	1.020			&	0.059			&	0.16			&	2.7			&	1580			\\
~~$\dot{P}_{mt,model}$ ($10^{-11}$ days/cycle)	&	0.11			&	0.037			&	0.051			&	0.17			&	0.020			&	...			\\
~~$\dot{P}_{mb}$ ($10^{-11}$ days/cycle)	&	-0.65			&	-0.080			&	-0.078			&	-0.51			&	-0.027			&	...			\\
~~$\dot{P}_{\Delta P}$ ($10^{-11}$ days/cycle)	&	-33.9			&	-14.8			&	-0.001			&	0.60			&	-57.7			&	-25000			\\
		\hline
	\end{tabular}
\end{table*}

In the second block of Table 5, I have collected a variety of the system properties.  Parameters for individual systems were collected from Schaefer et al. (2019), Szkody \& Ingram (1994), Gessner (1975), and Campbell \& Shafter (1995) for QZ Aur, and Salazar et al. (2017) for V1017 Sgr.  The light curve classes and magnetic natures come from Strope et al. (2010).  Distance, extinction ($A_V$), quiescent V magnitude ($V_q$), and quiescent absolute V magnitude ($M_{V,q}$) are taken from Schaefer (2018).  The accretion rate ($\dot{M}$ in units of M$_{\odot}$/year) is taken from an approximate relation $\dot{M}=10^{-0.4(M_{V,q}+16)}$ as taken from Dubus, Otulakowska-Hypka \& Lasota (2018), and also as $\dot{M}=2\times 10^{-11} (P/1hr)^{3.2}$ (Rappaport, Verbunt, \& Joss 1983).  I have included another estimate of the accretion rate, this one for the very long term average ($\dot{M}_{mb,model}$) as based on the model of Knigge et al. (2011).  The mass of the WD ($M_{WD}$) and mass of the companion star ($M_{comp}$) are from the catalog of Ritter \& Kolb (2003), with the mass ratio $q$=$M_{comp}$/$M_{WD}$ calculated.  The orbit's semi-major axis ($a$), the Roche lobe radius for the companion star ($R_{comp}$), the orbital velocity of the WD ($V_{WD}$), and the orbital velocity of the companion ($V_{comp}$) are from the usual Kepler's law equations in Frank, King, \& Raines (2002).  The expansion velocity of the nova ejecta is characterized by the expansion velocity of the nova shell long after the eruption (Downes \& Duerbeck 2000) and the width of the emission lines (Payne-Gaposhkin 1964).  No expansion velocity information is known from the eruptions of QZ Aur and V1017 Sgr, so I adopt a typical value of 1000 km s$^{-1}$.  $M_{ejecta}$ can be estimated from $M_{WD}$ and $\dot{M}$ as interpolated in the tables of Yaron et al. (2005).  (Both theoretical and observational estimates of $M_{ejecta}$ have real total uncertainties of over two orders-of-magnitude, so the tabulated values can only be regarded as relative values of bad accuracy.)  The similar trigger mass ($M_{ignition}$) is from the $M_{WD}$ and $\dot{M}$ values applied to Fig. 1 of Townsley \& Bildsten (2005).  The estimated recurrence time between nova eruptions ($\tau_{rec}$) is an average of the value in Yaron et al (2005), $M_{ejecta}/\dot{M}$, and $M_{ignition}/\dot{M}$.  The surface temperature of the companion star ($T_{comp}$) is variously taken from a blackbody fit to the spectral energy distribution, the spectral classification from absorption lines, and the relation with the orbital period (equation 14 in Patterson 1984).  The surface temperature, radius, and mass of the companion give the atmospheric scale height, $H$.

One use of these collected system properties is to see if the  sample of six CNe is extreme, biased, or special in any way.  For this, the answer is that the six appear to be a random collection of ordinary CN systems.  We really expect a fair sampling of CNe, because the six were largely selected based on their appearances from Earth (distance, inclination, and year of eruption) rather than any intrinsic property.  The eruption light curves are S-, D-, J-, and F-classes, while the $t_3$ values range from 25--231 days.  The WD masses range from 0.6--1.1 M$_{\odot}$.  Some of the WDs are magnetic (DQ Her, BT Mon, and RR Pic), while the others are not.  The mass ratios go from 0.4 to 0.95.  The periods vary similar to the distribution for CNe, while V1017 Sgr does have the longest period CN, RR Pic is near the shortest period for a CN.  The point is that these six CNe are fully representative of all CNe, so conclusions for these six can be extended to cover all CNe.

Another use of these system properties is to seek any correlation with $\Delta$P and $\dot{P}$.  Unfortunately, I have found no non-trivial correlation that looks to be significant.  For example, BT Mon is the only system with a positive $\Delta$P/P and also the system with the highest measured $V_{ejecta}$, but RR Pic has by-far the most-negative $\Delta$P/P but only has a middling $V_{ejecta}$.  For another unconvincing example, HR Del and RR Pic both have a positive $\dot{P}$ and have high values of the observed $\dot{M}$, but DQ Her has a zero $\dot{P}$ and the lowest $\dot{M}$, while QZ Aur has a far negative $\dot{P}$ yet a middling $\dot{M}$.  In all, I can find no useful or convincing correlation.

Another use of these system properties is to calculate predicted properties for the Hibernation model and the MBM.  The third block of Table 5 contains observed and calculated values of $\Delta$P/P in ppm, all for testing the Hibernation model.  The first line of this third block just repeats the observed value from the first block for ease of comparisons.  The second line gives $\Delta$P$_{ml}$/P, for the calculated effect due to the mass loss by the WD during the nova event.  The third line give the calculated $\Delta$P$_{FAML}$/P to give the predicted effects on the period change due to the FAML angular momentum loss during the eruption.  The fourth line gives the calculated $\Delta$P$_{jet}$/P for $\xi$ equal to +1 and -1, which corresponds to something near the maximum feasible effect of asymmetric ejection of the shell.  The fifth line gives the minimum $\Delta$P$_{5mag}$/P for which the quiescent nova fades by 5 mags, corresponding to the least value that anyone would care to call `hibernation'.  The sixth line gives the minimum $\Delta$P$_{M_q=+12}$/P for the quiescent system to go to hibernation with an absolute magnitude of +12.

A problem apparent from the third block is that four of the novae (QZ Aur, HR Del, RR Pic, and V1017 Sgr) have the maximum $\Delta$P/P for jetting as being smaller than the observed $\Delta$P/P.  That is, on the face of it, none of the physical mechanisms can account for the large observed $\Delta$P values.  For QZ Aur as an example, the mass-loss and FAML effects are small, while $\Delta$P$_{jet}$/P can vary from near -42 (for $\xi$=+1) to +42 (for $\xi$=-1) ppm, but there is no way to add the effects to get to the observed value of -290.71 ppm.  Having $\xi$ values near $\pm$1 is apparently common enough, and a narrow jet of ejecta can even increase the range of $\xi$ to -2 to +2, but this is still not enough to get the large observed value.  For most of the CNe, the excess factor is not large, but for RR Pic we need to get to the observed value of -2005 ppm.  So how can we physically get such large $\Delta$P values?  I think that the easiest explanation is simply to realize that the theoretical estimates of $M_{ejecta}$ used for the calculation has real uncertainties of orders-of-magnitude.  (Appendix A of Schaefer 2011 documents the large inconsistencies and circularities for one nova, and these are typical of all novae.)  So it is easy to allow the $M_{ejecta}$ to be ten-times larger than in the model of Yaron et al. (2005) for QZ Aur, all with the ordinary value of $\xi$=+0.7.  Still, for RR Pic with the fractional period change near 0.2 per cent, it will be hard to arrange for the huge loss of angular momentum across the eruption.  With the chaotic mixing of fast and slow gas in complex magnetic fields with binary motions, I can easily imagine that additional angular momentum loss mechanisms can operate briefly during the eruption.  Speculatively, I can wonder whether the high magnetic field of the WD in the RR Pic system can somehow lose its rotational angular momentum to the ejecting gas of the nova shell (i.e., magnetic braking of the WD), but this idea has the larger problem of how the lost angular momentum can be transferred from the WD rotation to the orbit.  Still, it is poor to think that a new physical mechanism, previously unsuspected, is required.  So I view the need to get an explanation for the large $\Delta$P values as being an open question, with a variety of possible answers.

The fourth block of Table 5 contains observed and calculated values for $\dot{P}$, all in units of 10$^{-11}$ days/cycle, all for testing the MBM.  The first line simply duplicates the observed $\dot{P}$ from the first block, for ease of comparison.  The second line gives the period change as taken from the best MBM model of Knigge et al. (2011), specifically the red curve in their Fig. 11, with this including both the effects of magnetic braking and mass transfer.  The third line give the $\dot{P}_{mt}$ value for the effect of steady mass transfer at the rate of the observed $\dot{M}$.  The next line gives $\dot{P}_{mt,model}$ for the assumed very-long-term average value of $\dot{M}_{mb,model}$ as taken from the MBM in Knigge et al. (2011).  The fifth line gives $\dot{P}_{mb}$, which is taken as $\dot{P}_{model}-\dot{P}_{mt,model}$.  The last line gives $\dot{P}_{\Delta P}$ for the observed sudden $\Delta$P as averaged out over the eruption cycle.

\section{Hibernation Model}

The fundamental basis for the Hibernation model is that the orbital period increases suddenly across a nova eruption, the stars separate in their orbits, and the accretion rate drops greatly.  That is, the Hibernation model requires $\Delta$P$>$0.  If $\Delta$P$<$0, then Hibernation cannot operate, and Hibernation is not operating.

Now, with my $\Delta$P program, we see that five-out-of-six CNe have negative period changes.  As these CNe are a representative sample of all CNe, we see that Hibernation is at-best uncommon amongst novae.  That is, we have proven that Hibernation is not working most of the time.  And the entire motivation of Hibernation and all of its utility, only comes around if most of the CNe undergo hibernation.  This is essentially a refutation of the entire Hibernation model.

We see the startling result that $\Delta$P/P must be greater than roughly +1000 ppm for even shallow hibernation to result.  Even for the one case of BT Mon with the observed $\Delta$P/P=+39.6$\pm$0.5 ppm, for even shallow hibernation, the period change would have to be 40$\times$ larger.  If we want real Hibernation where the accretion turns off (with $M_{V,hib}$$>$+12), we must have $\Delta$P/P$>$+2500.  So even though BT Mon has a positive period change across its eruption, it cannot possibly have separated by enough to cause hibernation in any sense.  Indeed, the brightness drop caused by the observed period change is so small as to be unobservable.  With this, we see that six-out-of-six CNe are certainly not behaving as required by Hibernation.  Going to six-out-of-six CNe, we have a much stricter version of the refutation of the Hibernation model.  

For all six of the CNe, the Hibernation model requires such a huge period change that there is no conceived explanation for how the binary can possibly separate enough so that even shallow Hibernation can occur.  Even with a narrow jetting of the entire nova shell in the reverse direction ($\xi$=-2), there is still no way for the stars to separate enough to be called hibernation.  That is, without inventing some new physical mechanism with an astoundingly large and positive size, Hibernation cannot ever work.

So, we have just shown that Hibernation is {\it not operating}, and the required period changes are larger than physically possible so Hibernation {\it cannot operate}.  This is a complete and utter refutation of Hibernation.  In hindsight, these simple and sure calculations should have been made back in the 1980s, but I know of no such results by anyone.

\section{Magnetic Braking Model}

The MBM predicts a particular value of $\dot{P}$ as a function of $P$, for example with the `standard model' of Knigge et al. (2011).  So now we have a distinct prediction that can be tested with my $\dot{P}$ data.  Well, V1017 Sgr has such a long orbital period that the magnetic braking mechanism certainly is not working as quantified by the model, so I only have five CNe with useful measured $\dot{P}$.  The fourth block in Table 5 collects the observational and theoretical values of $\dot{P}$.  The testing of the model will come from the various comparisons between the lines. 

The first comparison that I would like to highlight is between $\dot{P}_{\Delta P}$ and $\dot{P}_{model}$.  The ratio $\dot{P}_{\Delta P}$/$\dot{P}_{model}$ is +63, +350, +0.04, -1.8, and +7900 for the five CNe in order from Table 5.  For QZ Aur, HR Del, and RR Pic, the magnetic braking effects are negligibly small.  For BT Mon, the total effect ($\dot{P}_{\Delta P}$+$\dot{P}_{model}$) has the opposite sign as the model effect alone, so the total effect will result in the model giving greatly wrong predictions.  Only for DQ Her is the magnetic braking dominant, where the model would give correct predictions.  With the MBM failing for 4-out-of-5 CNe, we have a serious challenge for the venerable model.

The second comparison is between the observed $\dot{P}$ and $\dot{P}_{model}$.  The MBM would have $\dot{P}$/$\dot{P}_{model}$ = +1.0.  From Table 5, the ratio $\dot{P}$/$\dot{P}_{model}$ equals +5.3, $-$94, 0.00, +6.9, and $-$190 for the five CNe.  That is, the Magnetic Braking Model never makes a reasonable prediction as for the observed $\dot{P}$.  Indeed, for 3-out-of-5 CNe, the observed $\dot{P}$ is not even substantially negative.  That is, 60 per cent of a fair sample of CNe are not even with the period steadily decreasing, with this being a requirement and hallmark of the Model.  This is a separate and serious challenge to the MBM.

On the face of it, both `serious challenges' are refutations of the MBM.  That is, all five of the CNe that can be used to test the MBM have turned out greatly failed predictions for the MBM.  These failures are for large deviations that are highly significant from robust measures, and we are now out of the small-number-statistics regime.  These problems are only just now apparent with my $\Delta$P program, because no one has previously looked at CNe in this way.  

Well, the venerable model has nice successful predictions (the period gap, the bounce period, and the general decline of $\dot{M}$ as $P$ declines), so I think that the new data only impeach some part of the MBM.  In particular, to retain the successes, we should retain the general AML scenario and adopt the level of $\dot{J}$ as found by Knigge et al. (2011).  But the details of the physical mechanism must change substantially in light of the new measures.  The particular mechanism of magnetic braking is certainly still operating (at some level of $\dot{J}$ that is uncertain by orders-of-magnitude).  So there must be additional effects that need to be added to the original venerable model, and these additional mechanisms are dominating over the standard magnetic braking mechanism.

I can think of means by which the basic AML scenario can be retained for each of the two challenges.  My two solutions to the two challenges are both assuming that the observed effects will average out to near zero when considered over very long evolutionary time scales.  These are not elegant solutions.  And importantly, there is no evidence to support my solutions.  Nevertheless, my evidenceless speculation is not-unreasonable, and some such solutions must be found.

The first challenge is that $\dot{P}_{\Delta P}$ dominates over $\dot{P}_{model}$ by large factors for the majority of CNe.  My solution is to require and to assume that the very long term average period change, $\langle \langle \dot{P}_{\Delta P} \rangle \rangle$, is close to zero in comparison with $\dot{P}_{model}$.  With this, CV evolution would consist of $P$ having large jerks up and down over time, as eruption after eruption passes, yet with the overall trend falling similar to the MBM model.  So my solution assumes that $\Delta$P varies greatly over a huge range, both positive and negative.  

We must have some sort of large variation in $\Delta$P values for each individual CN.  That is, if each CN has an essentially constant $\Delta$P eruption after eruption, then the $\dot{P}_{\Delta P}$ would quickly drive the evolution to completion.  For example, with RR Pic losing 0.2 per cent of its period each eruption, then the time scale for going to zero-$P$ is 500 eruptions or 100,000 years.  The period obviously will not be driven to zero, but the simple time scale shows an evolution much too fast for the number of such novae discovered.  A more meaningful time scale is that RR Pic would be driven from its current period to the top edge of the period gap in roughly 50 eruptions over 10,000 years.  Similar considerations show that all the other CNe (except DQ Her) will be evolving too fast if their observed $\Delta$P holds steady eruption-to-eruption.  This means that CNe must have their $\Delta$P values changing greatly from nova to nova.

It is easy to imagine scenarios where $\langle \langle \dot{P}_{\Delta P} \rangle \rangle$$\approx$0, for example if the wide-and-weak jets of asymmetric shell ejection are random in direction, then the kicks should average out to near zero.  A potential problem arises because the jerks high and low are so large (up to a factor of 7900 large) that many eruptions are needed to average out random jerks.  From above, a typical value of $\dot{P}_{\Delta P}$/$\dot{P}_{model}$ is of order 100.  So it would take $\sim$10,000 eruptions to lower the average $\Delta$P jerks to being comparable to the size of $\dot{P}_{model}$.  For $\tau_{rec}$ or 1,000 or 100,000 years (Yaron et al. 2005), the required time interval to produce a small-enough time average will be of order 10--1000 million years.  This might be short enough of an interval so as to average out for evolutionary purposes.  Certainly, a detailed numerical model is required to resolve these issues.  Further, some physical understanding is needed as to how all the CNe have managed to fine tune their $\langle \langle \dot{P}_{\Delta P} \rangle \rangle$ to so close to zero.

For some CVs, the $\dot{P}_{\Delta P}$ values might be small because the nova recurrence time scale becomes very long.  For CVs above the period gap with $M_V,q$ around +9, the recurrence time scale becomes a million years.  Or maybe some of those systems avoid a nova eruption completely.  The CNe used in this study all have luminous $M_V,q$, so they are selected for short $\tau _{rec}$ by being classical novae.  But roughly 14 per cent of discovered CNe (the V1500 Cyg novae) have $M_V,q$ near +9 (Schaefer \& Collazzi 2010; Schaefer 2018), so it is clear that CNe do go to very long $\tau_{rec}$.  This might be a partial solution for how to minimize $\langle \langle \dot{P}_{\Delta P} \rangle \rangle$.

In all, if the $\Delta$P effects somehow manage to average out on evolutionary time scales, then we can get back to the MBM case, at least as a very jerky average.  But not really.  The trouble is that the $\Delta$P effects are usually dominant over magnetic braking, so the degree of connectedness for the binary will vary from MBM greatly after each and every eruption.  This will greatly change the demographics derived.  What we need is a real physical model for the processes that make for the huge variations in both $\dot{P}$ and $\dot{P}_{\Delta P}$ over long times, and these physical effects must be added in to the current MBM.  With $\Delta$P dominating, we might want to label the improved physical model by some name that does not include "magnetic braking".

Let me now point out some of the implications of the large $\Delta$P effects, if they vary randomly.    Start with a CV that happens to have the average $\dot{M}_{model}$.  If in one particular eruption, $\Delta$P is fairly large and positive, then the system will lower its $\dot{M}$ in response.  If a few eruptions in succession happen to have large positive $\Delta$P, then the system will become a dwarf nova, and its nova events will be very rare with a long $\tau_{rec}$. If the system randomly has a large negative $\Delta$P, or several such in a row, then the system will become a high-$\dot{M}$ classical nova, erupting with a fairly short $\tau_{rec}$.  Within this schematic middle-term evolutionary model, the $\Delta$P jerks will make the system random-walk between high-$\dot{M}$ CNe and low-$\dot{M}$ dwarf novae.  That is, the eruption-to-eruption variability of $\Delta$P could be what drives the wide observed deviations from $\Dot{M}_{model}$ for a given $P$.

My second serious challenge to MBM is that my observed $\dot{P}$ values never agree with the MBM prediction, indeed with 60 per cent not even having the hallmark and required negative value.  Well, the observed values are just snapshots from some century-long time interval, whereas the model predictions are for averaging over perhaps a million years.  It is conceivable that the period changes vary on time scales from 100--1,000,000 years, and that the very long term average, $\langle \langle \dot{P} \rangle \rangle$, somehow manages to equal $\dot{P}_{model}$.  In this case, the steady period changes will be waving up and down by large amounts on long time scales, but these large amplitude wavings all cancel each other out.  This evidenceless speculation has no precedent and no understanding.  But something like this is needed if the AML aspect of MBM is to survive.

I am reporting a new type of measures for CNe ($\Delta$P and $\dot{P}$ values for six systems) such as no one has looked before, and I find that the MBM predictions all fail horribly.  But the basic AML scenario really has to be correct.  I am not claiming that the MBM is wrong, rather that it is incomplete, requiring major new effects to be added in.  The MBM model has not included the effects of a previously unknown mechanism (whatever is making the $\Delta$P), and this mechanism is dominant over magnetic braking, and the MBM has not included whatever mechanism is making for the observed steady $\dot{P}$ between eruptions to vary so widely, both positive and negative.  So for any realistic CV evolutionary model in the future, it must include the physical effects of the $\Delta$P during each eruption, as well as the necessary large variations in the steady $\dot{P}$ between eruptions..

\section{Conclusions}

This paper presents the final results of the CN part of my 36-year-long $\Delta$P program.  For period changes across CN eruptions, I had previously reported the measure of $\Delta$P for BT Mon (Schaefer \& Patterson 1983), V1017 Sgr (Salazar et al. 2017), and QZ Aur (Schaefer et al. 2019).  In a companion paper, I report my measure of $\Delta$P for DQ Her plus a remeasure for BT Mon.  This paper gives the results for RR Pic and HR Del.  Here are my conclusions:

(1) I have measured six sudden orbital period changes across nova eruptions.  I find that the fractional period changes ($\Delta$P/P in ppm) are $-$290.71$\pm$0.28 for QZ Aur, $-$472.1$\pm$4.8 for HR Del, $-$4.46$\pm$0.03 for DQ Her, +39.6$\pm$0.5 for BT Mon, $-$2003.7$\pm$0.9 for RR Pic, and $-$273$\pm$61 for V1017 Sgr.  Five of six have negative signs, showing that the orbital period {\it decreased} across the nova event.

(2) I have the first measures of the steady period change during the quiescence ($\dot{P}$) of any classical novae.  I measure that the $\dot{P}$ parabolic terms (in units of $10^{-11}$ days/cycle) are $-$2.84$\pm$0.22 for QZ Aur, +4.0$\pm$0.9 for HR Del, 0.00$\pm$0.02 for DQ Her, $-$2.3$\pm$0.1 for BT Mon, +1.25$\pm$0.01 for RR Pic, while the uncertainty is too large to be useful for V1017 Sgr.  Only two of five have negative signs, showing that the majority of orbits that are {\it not} grinding down to short periods.

(3) Five out of my six CNe have negative-$\Delta$P, so Hibernation is certainly not working for these.  And with the realization that the positive-$\Delta$P for BT Mon is 40$\times$ too small to allow for any version of Hibernation, we really have six-out-of-six CNe being complete failures for the requirements of Hibernation.  These six CNe are a representative sample of all CNe, so we know that Hibernation can at most operate rarely.  Even if Hibernation operates rarely, then the model does not describe CV evolution or demographics.  This is a confident observational refutation of Hibernation.

(4) Detailed calculations show that for Hibernation to work (i.e., by dropping the accretion rate down so that it has nearly stopped, with $M_{V,hib}>$+12), the $\Delta$P/P values must be $>$+1300 ppm.  Further, detailed calculations of all known mechanisms (mass loss, FAML, the magnetic braking in the nova wind, and the newly invented jetting) show that they all individually (or in sum) are too small, by far, to possibly make for such large required period changes.  That is, there is no possibility that Hibernation could ever get the accretion to drop by enough to be noticeable.  This is a confident theoretical refutation of Hibernation.

(5) The confidently observed effects of the sudden period change caused by the nova, as spread out along the entire eruption cycle, make for a time-averaged period-change ($\dot{P}_{\Delta P}$) that is  much larger in amplitude than the standard period change expected for MBM in 80 per cent of the CNe.  In particular, the ratio $\dot{P}_{\Delta P}$/$\dot{P}_{model}$ has measured values of +63, +350, +0.04, $-$1.8, and +7900.  That is, in most cases, the effects of magnetic braking are negligibly small by factors of typically 100$\times$.  This is a serious challenge to the MBM.

(6) The hallmark and requirement of the MBM is that $P$ decreases over time.  The MBM predicts and requires that $P$ suffer a secular decrease, leading to a universal relation giving $\dot{P}$ as a function of $P$, at least as a long-term average.  The model requires that $\dot{P}$/$\dot{P}_{model}$ equal +1.0, whereas the five observed values are +5.3, -94, 0.00, +6.9, and $-$190.  So the MBM never makes a reasonable prediction for the observed $\dot{P}$.  Indeed, 60 per cent of the measures do not even have a decreasing $P$.  This is the second and independent serious challenge to the MBM.

(7) So, looking at CNe for the first time by means of measuring $\Delta$P and $\dot{P}$, I find that the MBM predictions have utterly failed in all cases.  But the venerable MBM has scored several nice predictions (the general decrease of $P$, the existence of the period gap, and the existence of the bounce period), so we know that the basic AML scenario is correct and we can reasonably estimate the total $\dot{J}$.  Hopefully and presumably, when the effects of the large-and-variable changes in $\Delta$P are averaged over many nova eruptions and when the effects of the large-and-variable $\dot{P}$ are averaged over many millennia, the long term averages will produce the correct level of angular momentum loss.  But the basic magnetic braking mechanism is usually negligibly small compared to the $\Delta$P and $\dot{P}$ effects.  So the future of CV evolution models is now keyed to advances in understanding and modeling the $\Delta$P and $\dot{P}$.  One exciting implication is the possibility that the wide range of $\dot{M}$ for a given $P$ could be driven as a random-walk of the $\Delta$P effects changing greatly from eruption-to-eruption. 

(8) The original motivation was to use the measured $\Delta$P values to derive $M_{ejecta}$ by a simple timing experiment with a confident dynamical basis.  ($M_{ejecta}$ is a critical quantity for many nova questions, but all previous measures, both observational and theoretical, are hopelessly poor with 2-or-more orders-of-magnitude uncertainty.)  Unfortunately, the existence of many $\Delta$P$<$0 values demonstrates that some other physical mechanism greatly dominates over effects from the mass loss to the system.  This means that the $\Delta$P values have no utility for measuring $M_{ejecta}$.  So our community is back to having no means of measuring $M_{ejecta}$ to better than 2 orders-of-magnitude.

(9) For CNe, I have exhaustively searched archival data for any other system that might allow a measure of $P_{pre}$, and I conclude that the six CNe in Table 5 are a complete list of all possible measures.  So there is nothing more that anyone can do, at least until some suitable CN erupts in the future, and then wait many years to get an adequately accurate $P_{post}$.  So Table 5 is all that anyone can know about CN $\Delta$P measures for a long time to come.

\section*{Acknowledgements}

The AAVSO helped with data archives, finder charts, and APASS comparison star magnitudes.  Funding for APASS has been provided by the Robert Martin Ayers Sciences Fund.  We are grateful for the historical observations and conservation made for the archival plates at Harvard and Sonneberg by many workers over the last 130 years.  For the work on RR Pic and HR Del, we are thankful for the hospitality and help from Alison Doane, Josh Grindlay, Peter Kroll, and the AAVSO Headquarters.  Nye Evans helpfully forwarded the SuperWASP light curve for HR Del.  Adrian Galan and Corey Meyers assisted with the observing HR Del at HRPO.  Gordon Myers, Peter Nelson, and Franz-Josef Hambsch provided many long time series showing the sinusoidal modulation of RR Pic.  I am also very thankful for the many of observers recording time series of HR Del.


%%%%%%%%%%%%%%%%%%%%%%%%%%%%%%%%%%%%%%%%%%%%%%%%%%

% Don't change these lines
\bsp	% typesetting comment
\label{lastpage}
\end{document}